\documentclass[aps,prb,twocolumn,superscriptaddress,showpacs,floatfix,amsmath,amssymb]{revtex4-1}
\usepackage{graphicx,xcolor}
\usepackage{mathtools}
\usepackage{graphicx,xcolor}
\usepackage{mathtools}
	
\def\K{\mathbf{K}}

\def\Q{\mathbf{Q}}	
\def\k{\mathbf{k}}

\begin{document}

\title{Typical-medium, multiple-scattering theory for disordered systems with Anderson localization}

\author{H.\ Terletska}
\email{hterlets@umich.edu}
\affiliation{Department of Physics, University of Michigan, Ann Arbor, Michigan 48109, USA}
\affiliation{Department of Physics \& Astronomy, Louisiana State University, Baton Rouge, Louisiana 70803, USA}
\author{Y.\ Zhang}
\affiliation{Department of Physics \& Astronomy, Louisiana State University, Baton Rouge, Louisiana 70803, USA}
\affiliation{Center for Computation \& Technology, Louisiana State University, Baton Rouge, Louisiana 70803, USA}
\author{L.\ Chioncel}
\affiliation{Augsburg Center for Innovative Technologies, University
of Augsburg, D-86135 Augsburg, Germany}
\affiliation{Theoretical Physics III, Center for Electronic
Correlations and Magnetism, Institute of Physics, University of
Augsburg, D-86135 Augsburg, Germany}
\author{D.\ Vollhardt}
\affiliation{Theoretical Physics III, Center for Electronic
Correlations and Magnetism, Institute of Physics, University of
Augsburg, D-86135 Augsburg, Germany}
\author{M.\ Jarrell}
\affiliation{Department of Physics \& Astronomy, Louisiana State University, Baton Rouge, Louisiana 70803, USA}
\affiliation{Center for Computation \& Technology, Louisiana State University, Baton Rouge, Louisiana 70803, USA}

\date{\today}

\begin{abstract}
The typical medium dynamical cluster approximation (TMDCA) is reformulated in the language of multiple scattering theory to make possible first principles calculations of the electronic structure of substitutionally disordered alloys including the effect of Anderson localization. The TMDCA allows for a systematic inclusion of non-local multi-site correlations and at same time provides an order parameter, the typical density of states, for the Anderson localization transition. The relation between the dynamical cluster approximation and the multiple scattering theory is analyzed, and is illustrated for a tight-binding model.
\end{abstract}


\maketitle

\section{Introduction}
During the last 50 years the study of the electronic~\cite{soven_67} and phononic~\cite{tayl.67} properties of substitutionally disordered alloys has been a very active field of research~\cite{el.kr.74,ziman_79}. Among the many theoretical methods proposed, the Green's function approach has proved to be particularly useful and convenient for calculating various physical quantities~\cite{a_gonis_92}.  One of the most successful and comprehensive schemes for the computation of the ensemble-averaged Green's function is the Coherent Potential Approximation (CPA)~\cite{ve.ki.68,yo.mo.73}, see also Refs.~\onlinecite{el.kr.74,ziman_79}. Today it is known that the CPA provides the exact solution for non-interacting fermions with diagonal (local) disorder on any lattice in the limit $Z \to \infty$, where $Z$ is the coordination number, provided the appropriate quantum scaling of the hopping amplitude is employed \cite{vlaming92}.  It is also possible to investigate \emph{interacting} disordered electrons by combining the dynamical mean-field theory  (DMFT)~\cite{MV89a,Janis91,ge.ko.92,Jarrell.92,pr.ja.95,ge.ko.96,ko.vo.04} with the CPA~\cite{ja.vo.92,ul.ja.95,kake.02}.

Using the CPA, the single-particle excitations in quench-disordered~\cite{soven_67,tayl.67,ziman_79,a_gonis_92,ve.ki.68,yo.mo.73,el.kr.74,jani.89} systems can be computed. In crystalline materials, quenched disorder manifests itself as randomly embedded impurities or alloy disorder. Therefore the CPA is frequently employed in the calculation of the electronic structure of these systems. The CPA introduces an effective crystalline medium in which a spatially fluctuating random potential is replaced by a purely local, but energy-dependent potential.  The effective potential is determined in such a way that the configurationally averaged Green's function is equal to the Green's function  of the  effective  medium. The CPA was also reformulated in the framework of the multiple scattering theory~\cite{gyor.72} and combined with the Korringa-Kohn-Rostoker (KKR) basis~\cite{jo.ni.86,LSMS,vi.ab.01}  or linear muffin-tin orbital (LMTO) basis~\cite{si.go.93} sets.  It has been used to calculate bulk properties~\cite{faul.82}, thermodynamic properties~\cite{jo.pi.93,ko.ru.95,ru.ab.95}, the phase stability~\cite{gy.st.83,al.jo.95,ab.ru.93,vito.07}, magnetic properties~\cite{ak.de.93,tu.ku.94,ab.er.95}, the surface electronic structure~\cite{ku.tu.92,ma.go.92,ab.sk.93,vito.07}, segregation~\cite{ru.ab.94,pa.dr.93}, and other alloy characteristics.

For decades attempts have been made to overcome the main shortcomings of the CPA by incorporating the missing nonlocal physics, e.g., by the molecular CPA~\cite{gonis_mcpa,a_gonis_92} and the dynamical cluster approximation (DCA)~\cite{Hettler98,m_jarrell_01a} in model Hamiltonian calculations.
Regarding the extension of the CPA for electronic structure calculations the KKR-Non-Local Coherent Potential Approximation (KKR-NLCPA)~\cite{Rowlands_2003} and the KKR-DCA approach~\cite{Biava} were proposed, both implementing the DCA coarse graining~\cite{m_jarrell_01a}. 

While the DCA and the KKR-NLCPA are able to include some nonlocal effects, they cannot describe the divergent behavior at the Anderson localization transition~\cite{Anderson,ab.an.79,le.ra.85,kr.ki.93,ev.mi.08,Abrahams}. This is due to the fact that these effective medium theories employ the {\it arithmetically} averaged density of states (ADOS), which does not became critical at the Anderson transition, and hence cannot serve as an order parameter.  As pointed out by Anderson~\cite{Anderson} one should instead determine the most probable (``typical'') value of the local density of states (LDOS), which is given by the global maximum of the full probability distribution function of the LDOS. In the case of a disordered system near the localization transition the LDOS fluctuates strongly, such that the corresponding probability distribution function possesses long tails. Indeed it has been demonstrated that the probability distribution function of the LDOS has very different properties in the metallic and insulating phase, respectively.~\cite{sc.sc.10,PhysRevB.89.081107} In particular, for weak disorder when states are extended, the probability distribution is Gaussian. By contrast, for strong disorder the probability distribution is asymmetric and is given by a log-normal distribution as obtained using analytic, field-theoretical approaches~\cite{PhysRevLett.72.526} for one-dimensional systems, or numerically exact calculations for three-dimensional lattices~\cite{sc.sc.10,PhysRevLett.105.046403,by.ho.10_a,by.ho.10_b}; 
see also Ref.~\onlinecite{PhysRevB.89.081107}.
The typical value of the LDOS is then determined by the geometric average~\cite{Crow1988}.

In an attempt to develop an order parameter formalism for Anderson localization, Dobrosavljevi\'{c} and collaborators~\cite{Vlad2003,Vlad2010,Vlad2010_b},
formulated an effective mean-field theory, the typical medium theory (TMT), which accounts for such changes in the probability distribution function. In particular, the TMT uses the 
geometrically averaged, i.e., typical, DOS in its self-consistency approach.  The typical DOS vanishes at the localization transition.  Thus, it can serve as an order parameter for Anderson localization. Despite this success, the TMT suffers from some of the same drawbacks as the CPA, i.e., it is still a local theory and hence does not include the crucial nonlocal quantum backscattering effects. Therefore the TMT can only provide a qualitative description of the Anderson localization transition.

Recently, some of us proposed the typical-medium dynamical cluster approximation (TMDCA)~\cite{TMDCA}, which extends the single-site TMT to a finite cluster and thereby allows for a systematic inclusion of nonlocal multisite  correlations. We demonstrated that the TMDCA overcomes the shortcomings of the TMT and is able to qualitatively and quantitatively describe Anderson localization.  We also extended this method to models with off-diagonal disorder\cite{PhysRevB.90.094208}, multiband systems~\cite{y_zhang_15a}, and  interactions~\cite{Ekuma2015}.

Until now, the TMDCA has only been applied to model  Hamiltonians~\cite{TMDCA,y_zhang_15a,y_zhang_16}. To incorporate this formalism directly into first-principles methods, the TMDCA should be reformulated such that access to the Green's function is provided in a language appropriate for calculations within density functional theory (DFT). This is best realized in the framework of multiple scattering theory. Therefore the purpose of the present study is to fill this gap and extend the TMDCA to the multiple scattering formalism. The advantage of this approach is its ability to treat both diagonal and off-diagonal disorder while requiring only small matrices due to the fast convergence of the scattering operators in angular momentum space~\cite{Gonis1997}.   In addition, we outline an alternative approach to first-principles downfolding, model-based  approaches\cite{y_zhang_15a} which is  able to identify the Anderson localization transition in real materials.

In particular, we show how the KKR-NLCPA can be extended to incorporate the typical medium formalism and demonstrate this in the case of a tight-binding model. The discussion of a material-specific implementation and the inclusion of electronic interactions will be postponed to future work. Section~\ref{sec:model} describes the model Hamiltonian. Section~\ref{sec:formalism} is devoted to a detailed discussion of the approximations for the ensemble-averaged Green's function and, in particular, to a discussion of the formal equivalence of the DCA~\cite{m_jarrell_01a} and the KKR-NLCPA as implemented in Refs.~\onlinecite{Rowlands_2003,Biava}.  These approximations are then applied to the tight-binding model of Sec.~\ref{sec:model}, and the correspondence between physical quantities calculated in the DCA and KKR-NLCPA is discussed. Section~\ref{sec:KKR-TMDCA} addresses the typical medium formulation within the multiple scattering approach: namely, we discuss the general algorithm and the formulas for calculating the typical density of states. Numerical results for the tight-binding model are presented in  Section~\ref{sec:results}, and  Section~\ref{sec:conclusion} contains a conclusion of the paper.

\section{Model and Formalism}
\label{sec:model}
\subsection{Tight-binding Model}
We consider the Anderson model of non-interacting electrons on a cubic lattice subject to a random diagonal potential, described by the Hamiltonian 
\begin{equation} \label{eqn:model}
H=-\sum_{\langle i j \rangle}W_{ij}(c_{i}^{\dagger}c_{j}+h.c.)+\sum_{i}V_i n_{i}\,.
\end{equation}
Here the operator $c_{i}^\dagger$($c_{i}$) creates (annihilates) an electron on site $i$, and $n_{i} = c_{i}^\dagger c_{i}$ is the number operator. The first term describes the hopping of electrons between nearest-neighbor sites $\langle i,j\rangle$ with the tight-binding hopping amplitude $W_{ij}=W$. We set $4W = 1$ as the energy unit.  The disorder is modeled through the energies $V_i$ of the local orbitals which are taken to be independent quenched random variables distributed according to some specified  probability  distribution $P(V_i)$. In the following, for illustrative purposes, we use a binary alloy distribution which has a bimodal disorder distribution $P(V_i)=c_A\delta(V_i-V_A)+(1-c_A)\delta(V_i-V_B)$, corresponding to a crystal randomly composed of $A (B)$ atoms at energy $V_A$ $(V_B)$ with concentration $c_A$ $(c_B)$. 

\subsection{Formalism}
\label{sec:formalism}
The main difficulty in dealing with disordered systems is the absence of translational invariance.  In order to use  approaches like the DCA one therefore has to average the free energy and its functional derivatives  (such as the Green's functions) over the possible disorder configurations. This is often justified since many experimental measurements, like in transport, spectroscopy and optical probes, tend to average over relatively large volumes and, thereby, over many local disorder configurations. Since this approach is computationally expensive, especially for large systems, one employs effective medium theories. These approaches are based on the idea that a heterogeneous medium can be replaced by an effective equivalent homogeneous medium. The problem is then reduced to finding a suitable representation for measured quantities, which in many cases is not provided by the average but by the full probability distribution. 

In this paper we demonstrate explicitly that the DCA and the KKR-NLCPA, when applied to a disordered, non-interacting tight-binding Hamiltonian, are equivalent. This fact, although having been observed in numerical calculations~\cite{Rowlands_2006}, has not been proven up to now. One goal of this paper is therefore to provide such a proof.  Furthermore, the TMT has thus far not been formulated in the multiple scattering language. The latter provides a theoretical framework for the construction of the typical medium formalism within the DFT. In order to see how the ideas of the TMDCA can be incorporated into a multiple scattering formalism (MS-TMDCA), we first briefly summarize the equations and concepts of the DCA approach, followed by the KKR-NLCPA procedure. 

Before proceeding with the formalism, let us note the conceptual relation between the DCA and the KKR-NLCPA. The DCA method can be successfully applied to strongly correlated electrons on a lattice with or without disorder~\cite{Maier}, to account for non-local dynamic correlations that capture spin and charge fluctuations in addition to configurational fluctuations. It should be mentioned that the KKR-NLCPA only accounts for disordered, {\it non}-interacting systems, i.e., the self-energy is associated with non-local correlations only due to configurational fluctuations. In this way the KKR-NLCPA can be viewed as the non-interacting limit of the DCA~\cite{m_jarrell_01a}.

The derivations given below closely follow Ref.~\onlinecite{m_jarrell_01a}, but we shall not use diagrammatic perturbation theory. Instead we focus on the formal connection between the DCA and KKR-NLCPA. Namely, both employ the DCA mapping of a lattice with $N$ sites onto a periodic cluster of size $N_c= L_c^D$  ($L_c$ is the linear dimension of the cluster, and $D$ is the dimensionality of the system) embedded in a self-consistently determined host.  Conceptually, this may be performed by partitioning the lattice with $N$ sites into clusters containing $N_c$ sites.  In reciprocal space this is equivalent to the division of the Brillouin zone (BZ) of the underlying lattice into $N_c$ cells of size $(2\pi/L_c)^D$, centered at the reciprocal sub-lattice vectors ${\bf K}$.  The lattice momenta within a given cell are denoted by $\tilde{\bf k}$. In the self-consistency loop, the Green's function is coarse-grained (averaged) over the momenta $\tilde{\k}$ surrounding the cluster momentum $\bf K$. The clusters are subject to periodic boundary conditions, which allows one to use the usual lattice Fourier transform.  Averaged cluster quantities possess translation invariance and can be taken from the reciprocal space into the real space (or vice versa) using the Fourier transform. In this approximation, correlations within the cluster are treated accurately up to a range $\xi \leq L_c$, while the physics on longer length scales is described at the static mean-field level. In the limit $N_c=1$ the purely local CPA is recovered.  By increasing the cluster size, the DCA systematically interpolates between the single-site and the exact result while remaining in the thermodynamic limit.  Here we consider the simple model described by Eq.~(\ref{eqn:model}) in $3D$, with a cluster size $N_c=38$.

The notation in the present paper uses the following convention:
\begin{itemize}
\item an underscore denotes a matrix in the cluster real space, 
\item superscripts $I,J$ indicate specific elements of real-space matrices,
\item an overbar represents quantities of the effective medium and the course-grained lattice quantities,
\item an argument ``$V$'' denotes quantities calculated on the cluster for a particular disorder configuration, 
\item arguments $\K$ and $\K'$ indicate that the corresponding quantities are coarse-grained or calculated in the cluster reciprocal space,
\item a subscript $l$ denotes local quantities,
\item the description ``effective medium'' denotes the homogeneous, or translationally invariant, problem on the lattice or cluster.
\end{itemize}

We use the short-hand notation $\langle...\rangle=\int dV_i P(V_i) (...)$ for disorder averaging. However, it is important to note that it is not necessary, nor even desired, to generate all disorder configurations.  This would cause the algorithm to scale as $2^{N_c}$ for the binary alloy model described above.  Rather, in the disorder-averaging procedure we generate the configurations stochastically and assume that the average restores the full point and space group symmetry of the lattice.  Then, by averaging over these symmetries, we effectively generate more disorder configurations.  As we will see below, the resulting algorithm scales as $N_c^3$.

\subsubsection{Dynamical Cluster Approximation}
\subsubsection*{ A) DCA algorithm I: \\
Reciprocal-space self-energy formulation}
\label{subsubsec:DCA-algo}
\label{sec:DCA}

To solve the disorder problem defined by Eq.~(\ref{eqn:model}), we first use the Dynamical Cluster Approximation (DCA), an effective medium cluster approximation in which the random potential of the Hamiltonian of Eq.~(\ref{eqn:model}) is replaced by the effective medium, defined by, as yet unknown, homogeneous self-energy $\bar{\Sigma}({\k, \omega})$.  In local (single-site) approximations, such as the Coherent Potential Approximation (CPA), all non-local corrections are neglected and the self-energy is a local quantity, i.e., in the CPA we approximate the lattice self-energy by a local self-energy  $\bar{\Sigma}({\k, \omega}) = \bar{\Sigma}(\omega)$ with only a frequency dependence.  To include non-local correlation effects in the DCA, the effective-medium lattice self-energy is approximated by a constant within each DCA cell in momentum space \cite{Hettler98,Hettler00}, $\bar{\Sigma}({\k, \omega}) = \bar{\Sigma}({\K, \omega})$.  The corresponding effective-medium lattice Green's function is then given by 
\begin{equation}
G(\k,\omega)=\frac{1}{w-\epsilon(\k)+\mu-\bar{\Sigma}(\K,\omega)},
\label{G_lattice}
\end{equation}
where $\k=\K+\tilde{\k}$ is the lattice reciprocal space vector, $\K$ is the DCA cluster vector, $\tilde{\k}$ is cluster momenta within each DCA cell, $\epsilon(\k)=-2W\left(\cos k_x+\cos k_y +\cos k_z \right)$ is the band dispersion, and $\mu$ is the chemical potential which is set to zero for a particle-hole symmetric case. 

To determine the DCA effective-medium self-energy $\bar{\Sigma}(\K,\omega)$, we must solve a $N_c$-site  cluster problem.  Following Ref.~{\onlinecite{Jarrell.92}} the effective-medium cluster Green's function is given by
\begin{equation}
\bar{G}_{cl}(\K,\omega)=\bar{g}_{cl}(\K,\omega)+\bar{g}_{cl}(\K,\omega)\bar{\Delta}(\K,\omega)\bar{G}_{cl}(\K,\omega)
\label{eq:Gcl}\,.
\end{equation}
Here $\bar{\Delta}(\K,\omega)$ is the hybridization function (the effective-medium ``bath''), which is obtained from integrating out all but the cluster degrees of freedom, and which describes the coupling between the cluster and the rest of the effective medium. We also introduce $\bar{g}_{cl}$, which is an ``isolated'' (with no hybridization to the effective-medium bath) cluster Green's function, and is defined as
\begin{equation}
\bar{g}_{cl}(\K,\omega)=\frac{1}{\omega-\bar{\Sigma}(\K,\omega)-\bar{\epsilon}(\K)}\label{eq:gcl}\,.
\end{equation}
Here $\bar{\epsilon}(\K)$ is the coarse-grained dispersion averaged over $\tilde{\k}$ points within the cell centered on $\K$
\begin{equation}
\bar{\epsilon}(\K) = \frac{N_{c}}{N}\sum_{\tilde{\k}} \epsilon_{\K+\tilde{\k}}
\label{eq:epsb}
\end{equation}
For the CPA formalism, which corresponds to the $N_c=1$ limit of the DCA, all cluster quantities are independent of momentum $\K$, and $\bar{\epsilon}(\K)$ is equal to a constant which here we set to zero. 

These formulas may be used to define a DCA algorithm where the effective-medium self-energy  $\bar{\Sigma}(\K,\omega)$ is required to give an exact description of the original random medium.  Therefore, to determine the yet-unknown  $\bar{\Sigma} (\K,\omega)$ and $\bar{\Delta}(\K,\omega)$, we replace the effective-medium potential $\bar{\Sigma}(\K,\omega)$ on the cluster by the random  potential $V_I$. We then demand that upon averaging the scattering caused by the random potential $V_I$ vanishes identically in the effective medium. This construction defines the self-consistency condition which determines $\bar{\Sigma}(\K,\omega)$. 

As a first step, the cluster Green's function will be calculated.  We first Fourier transform the cluster Green's function of Eq.~(\ref{eq:Gcl}) into the cluster real space. This is simplified by the fact that the Green's function of the effective medium is translationally invariant. Hence, the Green's function between sites $(I,J)$ belonging to the cluster in the effective medium is
\begin{equation}
 \bar{G}^{IJ}_{cl}(\omega)= \bar{g}_{cl}^{IJ}(\omega)+\sum_{K,L} \bar{g}_{cl}^{IK}(\omega)\bar{\Delta}^{KL}(\omega)\bar{G}^{LJ}_{cl}(\omega)\,.
\label{eq:Gcluster}
\end{equation}
Equivalently, Eq.~(\ref{eq:gcl}) can be written (in real space) in the matrix form
\begin{equation}
 \bar{\underline{g}}_{cl}(\omega)=\left(\omega \mathbb{I} -\bar{\underline{\Sigma}}(\omega)-\bar{\underline{W}}) \right)^{-1}\,
\label{eq:gclrs}
\end{equation}
where $\bar{W}^{IJ}=\sum_{\K}\bar{\epsilon}(\K)e^{i\K(\bf{R}_I-\bf{R}_J)}$ is the Fourier transform of the cluster coarse-grained dispersion.

Now, to consider the real disorder cluster problem embedded in the effective medium, everywhere on the cluster we replace the effective potential $\underline{\bar\Sigma}(\omega)$ of Eq.~(\ref{eq:gclrs}) by the random potential $V_I$. The corresponding  disorder cluster Green's function is then given by
\begin{eqnarray}
G_{cl}^{I J}(\omega,V) &=& g_{cl}^{I J}(\omega,V) \\
&+& \sum_{K L}g_{cl}^{I K}(\omega,V)\bar{\Delta}^{K L}(\omega)G_{cl}^{K J}(\omega,V)\,,
\nonumber
\label{eq:-21}
\end{eqnarray}
where the isolated cluster Green's function $\underline{g}_{cl}(\omega,V)$ for a given disorder configuration $V$ is given by
\begin{equation}
\underline{g}_{cl}(\omega,V)=\left(\omega \mathbb{I}- \underline{V} - \underline{\bar W}\right)^{-1}\,.
\end{equation}
Since the hybridization function $\underline{\bar{\Delta}}(\omega)$ is disorder independent, it represents the same hybridization function of the effective medium of Eq.~(\ref{eq:Gcl}).  Notice, that the above Green's function can be rewritten in terms of the cluster-excluded (cavity) Green's function $ {\mathcal{G}}^{I J}(\omega)$, conventionally used in the DCA community~\cite{Jarrell.92,Maier}, as
\begin{equation}
\underline{G}_{cl}(\omega,V)=\left(\underline{{ \mathcal{G} } }^{-1}(\omega)- \underline{V}\right)^{-1} \,,
\label{eq:-26}
\end{equation}
with
$
\underline{{\mathcal{G}}}^{-1}(\omega)=\left(\omega \mathbb{I} - \underline{\bar W}-\underline{\bar \Delta}(\omega)\right)\,.
$
Due to this inverse, this algorithm and the subsequent algorithms below, scale with cluster size as $N_c^3$.

In the next step we impose the DCA self-consistency condition, which requires that the disorder-averaged cluster Green's function is equal to the effective medium cluster Green's function, with
\begin{equation}
\left< G^{IJ}_{cl}(\omega,V)\right> =\bar{G }^{IJ}_{cl}(\omega)\,.
\label{eq: -27}
\end{equation}
Here $\left< ...\right>$ denotes averaging over disorder configurations on the cluster. Or, equivalently in momentum space, we write it as
\begin{equation}
\bar{G}_{cl}(\K,\omega)=FT(\left <G^{IJ}_{cl}(\omega,V)\right >),
\end{equation}
where FT stands for the Fourier Transform of the disorder-averaged cluster Green's function to the cluster reciprocal space. This condition is then used to update the self-energy, with
\begin{equation} 
\bar{\Sigma}(\K,\omega) = \omega - \bar{\epsilon}(\K) - {\bar{\Delta}}(\K,\omega)  -\bar{G}_{cl}^{-1}(\K,\omega)\,,
\label{eq:sigmanew}
\end{equation}
which is then used to calculate the coarse-grained lattice Green's function $\bar{G}(\K,\omega)$. 
\begin{equation}
\bar{G}(\K,\omega)=\frac{N_{c}}{N}\sum_{\tilde{k}}\frac{1}{\omega+\mu-\epsilon_{\K+\tilde{\k}}-\bar{\Sigma}(\K,\omega)} \,.  
\label{eq:dca}
\end{equation}
At convergence this is identical to the effective medium cluster Green's function $G_{cl}(\K,\omega)$.
Given the self-energy $\bar\Sigma (\K,\omega)$ through Eq.~(\ref{eq:sigmanew}) one defines the new hybridization function $\bar{\Delta}(\K,\omega)$ as: 
\begin{equation}
{\bar{\Delta}}(\K,\omega) = \omega - \bar{\epsilon}(\K) - \bar{\Sigma}(\K,\omega) - \bar{G}^{-1}(\K,\omega)\,,
\label{eq:deltanew}
\end{equation}
which closes the DCA loop.

\begin{figure}[!t]
\includegraphics[trim = 0mm 0mm 0mm 0mm,width=\columnwidth,clip=true]{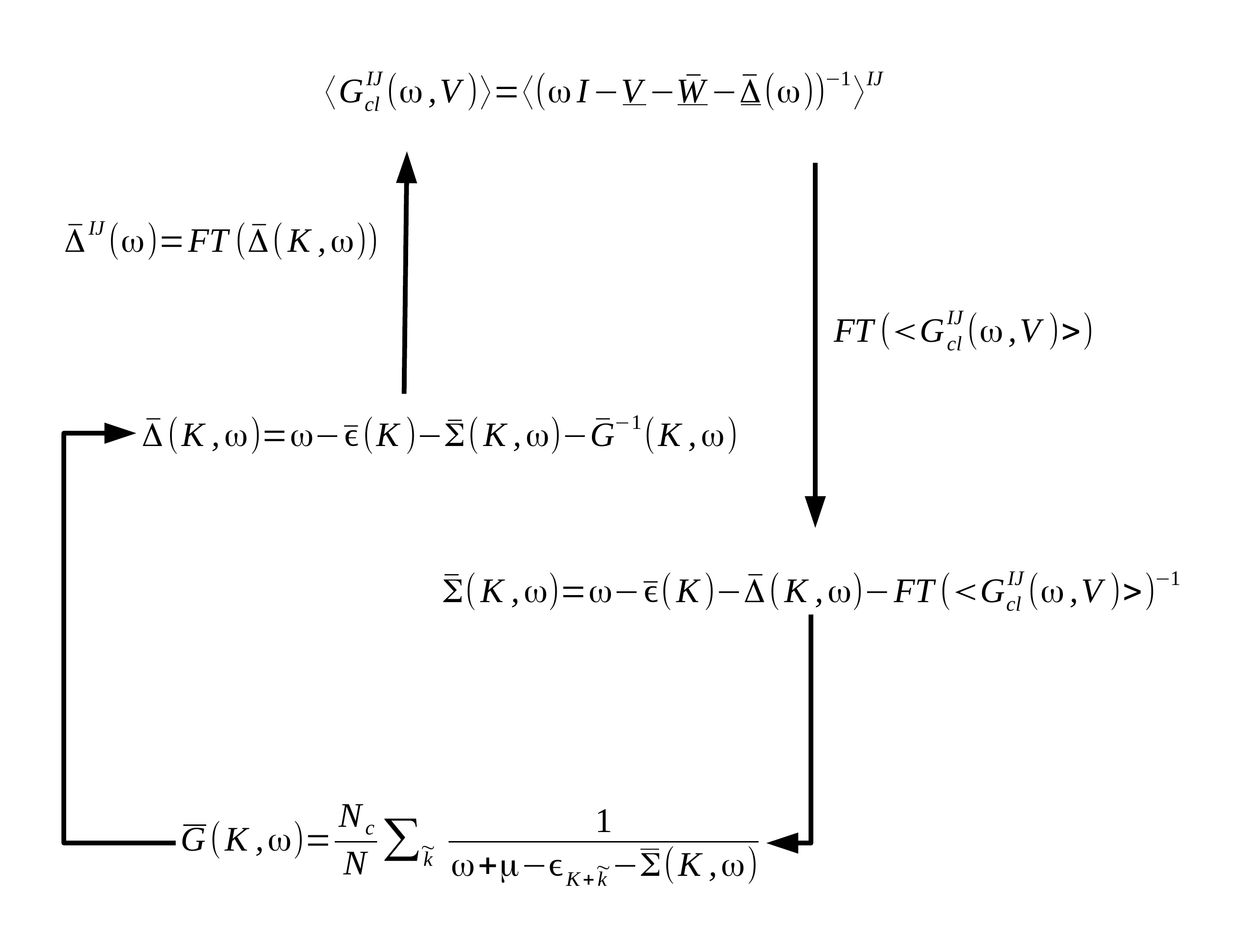}
\caption{The DCA self-consistency loop. The arrows correspond to the 
steps taken in the DCA algorithm described in Subsection~\ref{subsubsec:DCA-algo}.}
\label{fig:DCA_loop}
\end{figure}

The numerical self-consistency loop is diagrammatically shown in Fig.~\ref{fig:DCA_loop}, and below we describe the DCA iterative procedure:
\begin{itemize}
\item[1)] First, a guess for the cluster self-energy $\bar{\Sigma}(\K,\omega)$ is made (usually set to zero) and the lattice coarse-grained Green's function $\bar{G}(\K,\omega)$ is calculated using Eq.~(\ref{eq:dca}).
\item[2)] The effective-medium hybridization function $\bar{\Delta}(\K,\omega)$ is constructed by solving Eq.~(\ref{eq:Gcl}) and Eq. ~(\ref{eq:gcl}), i.e., $\bar{\Delta}(\K,\omega)=\omega-\bar{\epsilon}(\K)-\bar{\Sigma}(\K,\omega)-\bar{G}^{-1}(\K,\omega)$. 
Since the cluster problem is solved in real space, we Fourier transform the obtained hybridization $\bar{\Delta}(\K,\omega)$  to real space of the cluster.
\item[3)] Next, the cluster problem is solved in real space and 
the disorder-averaged cluster Green's function $\langle G_{cl}(\omega,V)\rangle^{IJ}=\langle(\omega \mathbb{I}-\underline{V}-\underline{\bar{W}}-\underline{\bar{\Delta}}(\omega))^{-1}\rangle$ is calculated.
\item[4)] Once the cluster problem is solved, we construct a new cluster self-energy, 
$\bar{\Sigma}(\K,\omega)=\omega-\bar{\epsilon}(\K)-\bar{\Delta}(\K,\omega)-\bar{G}_{cl}^{-1}(\K,\omega)$,
using the Fourier Transform of the disorder-averaged cluster Green's function to the cluster reciprocal space, i.e., $\bar{G}_{cl}(\K,\omega)=FT(\langle G_{cl}(\omega,V)^{IJ}\rangle)$.
\end{itemize}
The self-consistent procedure is repeated until $\bar{\Sigma}(\K,\omega)$ converges to the desired accuracy.

We note that an equivalent self-consistency loop can be constructed by using the cluster-excluded Green's function $ \underline{\mathcal{G}}(\omega)$.~\cite{m_jarrell_01a} This can be done by noting that $\bar{\Sigma}(\K,\omega) = {\mathcal{G}}^{-1}(\K,\omega) - \bar{G}_{cl}^{-1}(\K,\omega)$.

\subsubsection*{B) DCA algorithm II:\\ Local and non-local self-energy contributions}
\label{subsubsec:DCA2}

While in the DCA formalism the non-local contribution to the self-energy, obtained for $N_c>1$, is encoded explicitly in the $\bar{\Sigma}(\K,\omega)$, in the KKR-NLCPA formalism the cluster extensions involve a separate  analysis of the local and non-local contributions. To provide a better connection between these methods, in the following we present an alternative DCA self-consistency analysis which involves an explicit separation of the local and non-local components of the self-energy. 
To this end, we introduce 
\begin{equation}
\tilde {\alpha} (\K,\omega)=\bar{\Sigma}(\K,\omega)-\bar{\Sigma}_l(\omega),
\label{eq:alpha}
\end{equation}
which defines the non-local corrections to the self-energy  (the subscript $l$ is used to emphasize the local, momentum independent, quantity). For the CPA (i.e. in the $N_c$=1 limit) this non-local contribution vanishes since the self-energy is purely local $\bar{\Sigma}(\K, \omega)\rightarrow \bar{\Sigma}(\omega)$). Using this definition, we rewrite the effective-medium lattice Green's function of Eq.~(\ref{G_lattice}) as 
\begin{equation}
\bar{G}(\k,\omega)=\frac{1}{\bar{g}_l^{-1}(w)-\tilde {\alpha} (\K,\omega)+\mu-\epsilon(\tilde{\k}+\K)},
\label{G_lattice-nonlocal}
\end{equation}
where we introduce the locator Green's function $\bar{g}_l(\omega)$ \cite{a_gonis_92}, defined as
\begin{equation}
\bar{g}_l(\omega)=\frac{1}{\omega-\bar{\Sigma}_l(\omega)}\,.
\label{eq. gloc}
\end{equation}
Similarly, by applying the decomposition of the self-energy into local and non-local parts, with $\bar{\Sigma}(\K,\omega)=\tilde{\alpha}(\K)+\bar{\Sigma}_l(\omega)$, we rewrite the effective-medium cluster Green's function of Eq.~(\ref{eq:Gcl}) and Eq.~(\ref{eq:gcl}) as
\begin{equation}
\bar{G}_{cl}^{-1}(\K,\omega)=\bar{g}_{cl}^{-1}(\K,\omega)-\bar{\Delta}(\K,\omega),
\label{Glc_nonlocal}
\end{equation}
with the isolated cluster Green's function 
\begin{equation}
\bar{g}_{cl}^{-1}(\K,\omega)=\bar{g}_l^{-1}(\omega)-\tilde{\alpha}(\K,\omega)-\bar{\epsilon}(\K)\,.
\label{eq. Gcl_nonlocal}
\end{equation}

Next, to find the yet unknown $\tilde{\alpha}(\K,\omega)$ and $\bar{g}_l(\omega)$ we consider the disordered cluster embedded in such an effective medium. Given a disorder configuration $V_I$ we construct the corresponding cluster Green's function 
\begin{equation}
\underline{G}_{cl}(V,\omega)=\left ( \underline{g}_{l}^{-1}(V,\omega)-\bar{\underline{W}}-\bar{\underline{\Delta}}(\omega)  \right ) ^{-1},
\label{eq. Gv_nonlocal}
\end{equation}
where in Eq.~\ref{eq. gloc} we replaced the effective-medium potential $\bar{\Sigma}^{IJ}(\omega)$ with the random-disorder potential $V_I$ for the locator Green's function, i.e.
\begin{equation}
\underline{g}_{l}(V,\omega)=(\omega \mathbb{I}-\underline{V})^{-1}\,.
\label{eq: gl_V}
\end{equation}
We then impose the DCA self-consistency condition, requiring that the disorder-average cluster Green's function and the effective-medium cluster Green's function are equal, i.e.,
\begin{equation}
\left <G^{IJ}_{cl}(\omega,V)\right >=\bar{G }^{IJ}_{cl}(\omega)\,.
\label{eq: self-cons}
\end{equation}
This allows us to define a new cluster self-energy, as in Eq.~(\ref{eq:sigmanew}).
Finally, we complete the self-consistency loop by calculating the lattice coarse-grained Green's function
\begin{equation}
\bar{G}(\K,\omega)=\frac{N_c}{N}\sum_{\tilde{\k}}\frac{1}{\bar{g}_l^{-1}(w)-\tilde {\alpha} (\K,\omega)+\mu-\epsilon(\tilde{\k}+\K)}.
\label{eq: cg_nonlocal}
\end{equation}

This DCA-II algorithm is close in spirit to the KKR-NLCPA scheme with an explicit separation of the local and non-local contributions of self-energy.  It is, of course, equivalent to the DCA-I algorithm described in the previous subsection. 
In Fig.~\ref{fig:DCA_loop2} we present the corresponding self-consistent loop of the computational procedure of the DCA-II algorithm. The steps of the iterative procedure are described as follows: 

\begin{figure}[!t]
\includegraphics[trim = 0mm 0mm 0mm 0mm,width=\columnwidth,clip=true, angle =0]{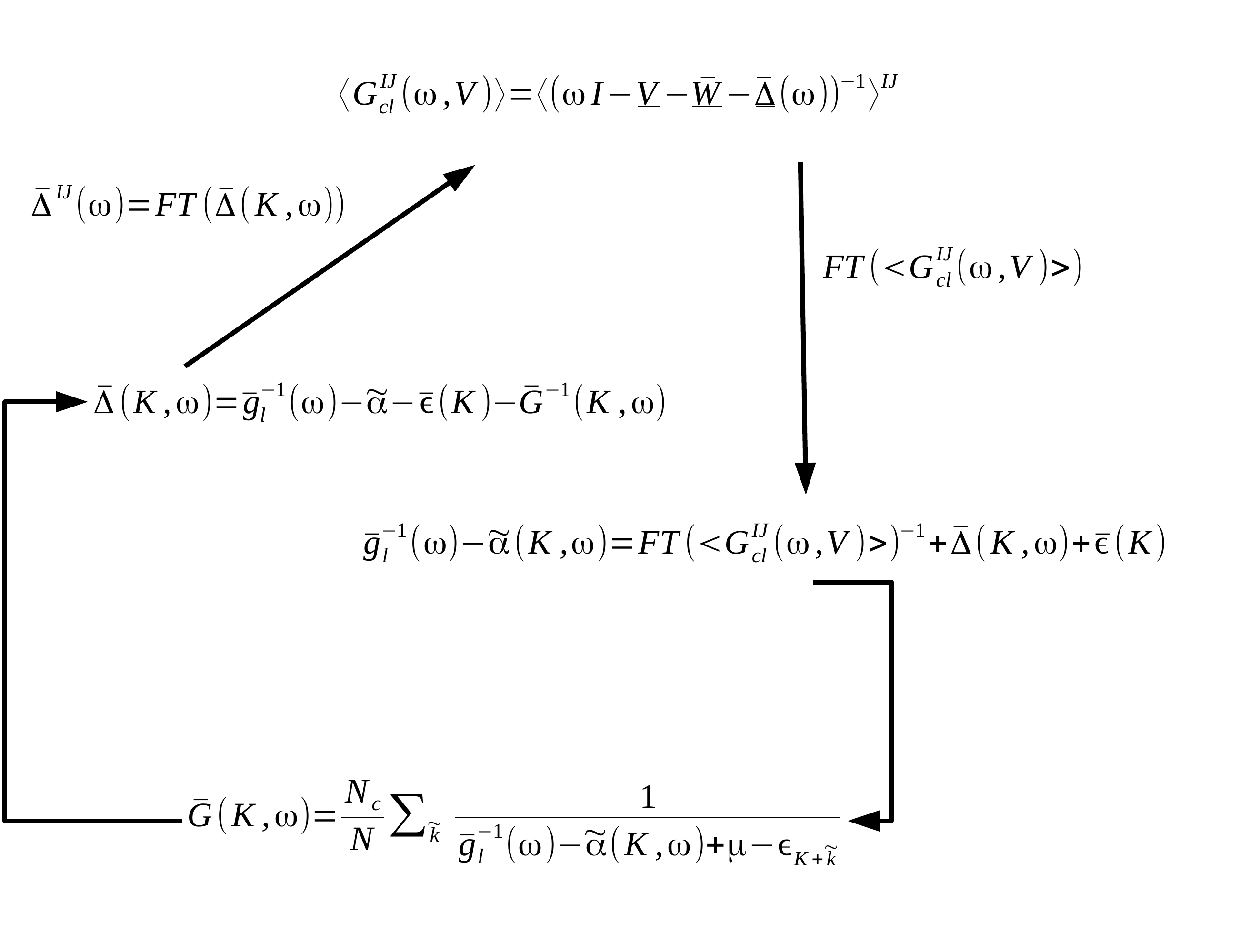}
\caption{An alternative DCA-II self-consistency algorithm with separate local and non-local contributions to the self-energy.}
\label{fig:DCA_loop2}
\end{figure}

\begin{itemize}
\item[1)] We compute the lattice coarse grained Green's function $\bar{G}(\K,\omega)$ using Eq.~(\ref{eq: cg_nonlocal}), by making a guess for $\bar{g}_l^{-1}(\omega)-\tilde{\alpha}(\omega)$ based on the values of the local and non-local self-energy components.  If nothing is known a priori, the guess $\bar{\Sigma}_l(\omega)=0$ and $\tilde{\alpha}=0$, with resulting $\bar{g}_l^{-1}(\omega)-\tilde{\alpha}=\omega$ may serve as the starting point. 
\item[2)] We then construct the effective-medium hybridization function $\bar{\Delta}(\K,\omega)$ by solving Eq.~(\ref{Glc_nonlocal}) and Eq.~(\ref{eq. Gcl_nonlocal}), i.e., $\bar{\Delta}(\K,\omega)=\bar{g}_l^{-1}(\omega)-\tilde{\alpha}(\K)-\bar{\epsilon}(\K)-\bar{G}^{-1}(\K,\omega)$. 
Since the cluster problem is solved in real space we Fourier transform the obtained hybridization function, $\bar{\Delta}(\K,\omega)$, to the real space cluster.
\item[3)] In the next step, we solve the cluster problem and calculate the disorder-averaged cluster Green's function $\langle\underline{G}_{cl}(\omega,V)\rangle=\langle( \omega \mathbb{I}-\underline{V}-\underline{\bar{W}}-\underline{\bar{\Delta}}(\omega))^{-1}\rangle$.
\item[4)] Once the cluster problem is solved, the disorder-averaged cluster Green's function $\langle\underline{G}_{cl} \rangle$ is used to construct an isolated cluster Green's function   $\bar{\underline{g}}_{cl}^{-1}=\langle\underline{G}_{cl}(V,\omega)\rangle^{-1}+\bar{\underline{\Delta}}(\omega)$, which we then use to get a new $\tilde{\underline{\alpha}}(\omega)=\bar{\underline{g}}_l^{-1}(\omega)-\bar{\underline{g}}_{cl}^{-1}-\bar{\underline{W}}$. Here $\bar{\underline{g}}_l$ is a local component of $\bar{\underline{g}}_{cl}$. 
Notice, that in practice we combine these two steps in momentum space and instead calculate $\bar{g}_l^{-1}(\omega)-\tilde{\alpha}(\K,\omega)=FT(\langle G_{cl}^{IJ}(V,\omega)\rangle)^{-1}+\bar{\Delta}(\K,\omega)+\bar{\epsilon}(\K)$,
where $\bar{G}_{cl}(\K,\omega)=FT(\langle G_{cl}^{IJ}(\omega,V)\rangle)$.
\item[5)]  We repeat the self-consistent procedure through steps 1-5 until convergence is obtained.
\end{itemize}

\subsection{KKR-Nonlocal Coherent Potential Approximation (KKR-NLCPA)}
\label{sec:KKR-tau}
\subsubsection{ KKR-NLCPA algorithm I }
\label{subsubsec:kkr1}

In this section we present the details of the KKR-NLCPA~\cite{Biava,Rowlands_2003} formalism applied to the tight-binding Hamiltonian. Essentially, the KKR-NLCPA is the static limit (i.e., with no inelastic scattering) of the DCA as derived by Jarrell and Krishnamurthy~\cite{m_jarrell_01a}. To demonstrate this, in this subsection we present the ``KKR-NLCPA algorithm I'' which is an alternative to Refs. ~\onlinecite{Biava,Rowlands_2003} in a spirit which is very similar to the original DCA scheme.

The derivation of the KKR-NLCPA makes use of the multiple scattering formulation in which the central quantity is the effective-medium cluster scattering path operator $ \bar {\tau} (\K,\omega)$ rather than the cluster effective-medium Green's function $\bar {G}(\K,\omega)$ used in the DCA.  The KKR-NLCPA is also an effective-medium method, where the original disorder problem of Eq.~(\ref{eqn:model}) is replaced by the effective-medium problem, such that the lattice effective scattering path operator $\bar{\tau}(\k,\omega)$ is given as  
\begin{equation}
\bar{\tau}(\k,\omega)=\frac{1}{\bar{t}^{-1}(\k,\omega)-G_0^{'}(\k,\omega)}\,.
\label{eq. tau_lattice}
\end{equation}
Here $\bar{t}^{-1}(\k,\omega)$ is the yet unknown homogeneous effective scattering $t$-matrix and $G_0^{'}(\k,\omega)$ is free space structure constant ~\cite{a_gonis_92}, which for tight-binding Hamiltonian corresponds to a bare Green's function $G_0(\k,\omega)$. 

To determine the $t$-matrix we use the DCA-like cluster embedding scheme, where the lattice t-matrix, is approximated by a cluster t-matrix with $\bar{t}(\k,\omega)=\bar{t}(\K,\omega)$ and is obtained by solving a cluster embedded in an effective-medium, with the  effective medium cluster scattering path operator defined as
\begin{equation}
\bar{\tau}_{cl}(\K,\omega)=\bar{t}_{cl}(\K,\omega)+\bar{t}_{cl}(\K,\omega)\bar{\Delta}^{'}(\K,\omega)\bar{\tau}_{cl}(\K,\omega)\,.
\label{eq:-tau}
\end{equation}
Here $\bar{\Delta}^{'}(\K,\omega)$ 
arises from integrating out all but the cluster degrees of freedom and 
corresponds to the hybridization function $\Delta(\K,\omega)$ in the DCA scheme.  In the KKR literature this quantity is referred as the ``effective-medium renormalized interactor'' (see for example Refs.~\onlinecite{ziman_79,a_gonis_92}). 

Here we also define $\bar{t}_{cl}(\K,\omega)$ the isolated cluster t-matrix (with no hybridization to the effective medium) as
\begin{equation}
\bar{t}_{cl}(\K,\omega)=\frac{1}{\bar{t}^{-1}(\K,\omega)-\bar G_{0}^{'}(\K,\omega)}
\label{eq:-tcl}\,,
\end{equation}
Due to its explicit $\K$-dependence, $t_{cl}(\K,\omega)$ takes into account non-local correlations up to the cluster size which are missing in the local KKR-CPA analysis for $N_c=1$. We note, that for $N_c=1$ this quantity becomes local with  $\bar{t}_{cl}({\K,\omega})\rightarrow \bar{t}_l(\omega)$, where the subscript $l$ indicates ``local quantity''.  Since the effective medium is translationally invariant, we Fourier transform  Eq.~(\ref{eq:-tau}) to the cluster real space, i.e.,
\begin{equation}
{\bar \tau}^{IJ}_{cl}(\omega)=\bar t_{cl}^{IJ}(\omega)+\sum_{K,L}\bar t_{cl}^{IK}(\omega)\bar{\Delta}^{'KL}(\omega)\bar{\tau}^{LJ}(\omega)\,.
\label{eq:-tau_rs}
\end{equation}
Equivalently, Eq.~(\ref{eq:-tcl}) in real space can be rewritten as
\begin{equation}
\bar{\underline t}_{cl}(\omega)=\left( \bar {\underline t}^{-1} (\omega)-\bar {\underline G}_{0}^{'}(\omega) \right )^{-1} \,.
\label{eq:tclrs}
\end{equation}

To determine the effective-medium quantities we next introduce the impurity
cluster, with the disorder placed on each cluster site, embedded in the effective medium. To do this, we replace the effective-medium quantities in Eq.~(\ref{eq:-tau_rs}) and Eq.~(\ref{eq:tclrs}) by their disorder-dependent counterparts. Note that $\underline{\bar\Delta}'(\omega)$ is independent of the disorder configuration of the cluster since it describes the effective medium. Hence it is not changed under the substitution of an effective cluster with the disorder-dependent cluster problem.  Thus, we obtain an expression for the cluster path operator for a given disorder configuration 
\begin{eqnarray}
\tau_{cl}^{I J}(\omega,V) &=& t_{cl}^{I J}(\omega,V) \\
&+&\sum_{K,L}t^{I K}_{cl}(\omega,V)\bar{\Delta}^{'K L}(\omega)\tau_{cl}^{L J}(\omega,V)\nonumber \,,
\label{eq:-10-1-1}
\end{eqnarray}
where the cluster $t$-matrix for a given disorder $V$ is 
\begin{equation}
{\underline t}_{cl}(\omega,V)=\left(  {\underline t}^{-1} (V)-\bar {\underline G}_{0}^{'}(\omega) \right )^{-1} ,
\label{eq:-10-1}
\end{equation}
where for the tight-binding model $\underline{t}^{-1}=\underline{V}^{-1}$.
With this transformation, the cluster path operator for a given disorder configuration reads
\begin{equation}
\underline{\tau}_{cl}(\omega,V)=\left(\underline{t}^{-1}(V)-\underline{\bar G}_{0}^{'}-\underline{\bar{\Delta}}^{'}(\omega)\right)^{-1}\,.
\label{eq:-10-1-1-1}
\end{equation}

In the next step, we impose the self-consistency condition, which requires that, when placing the cluster in the effective medium, no additional scattering is produced on average, i.e.,
\begin{equation}
\left<\tau_{cl}(\omega,V)\right>^{I J}=\bar\tau^{I J}_{cl}(\omega).\label{eq:-6}
\end{equation}
We then close the self-consistency loop by calculating the coarse-grained lattice scattering path operator given as
\begin{equation}
\bar{\tau}(\K, \omega)=\frac{N_c}{N}\sum_{\tilde{k}}\frac{1}{\bar{t}^{-1}(\K,\omega)-G_0^{'}(\K+\tilde{\k},\omega)},
\label{eq: coarse-grained tau}
\end{equation}
where $\bar{t}^{-1}(\K,\omega)=FT(\left<\tau_{cl}(\omega,V)\right>^{I J})+\bar G_{0}^{'}(\K,\omega)+\bar{\Delta}^{'}(\K,\omega)$ is obtained from Eq.~(\ref{eq:-tau}) and Eq.~(\ref{eq:-tcl}).

\begin{figure}[!t]
\includegraphics[trim = 0mm 0mm 0mm 0mm,width=\columnwidth,clip=true]{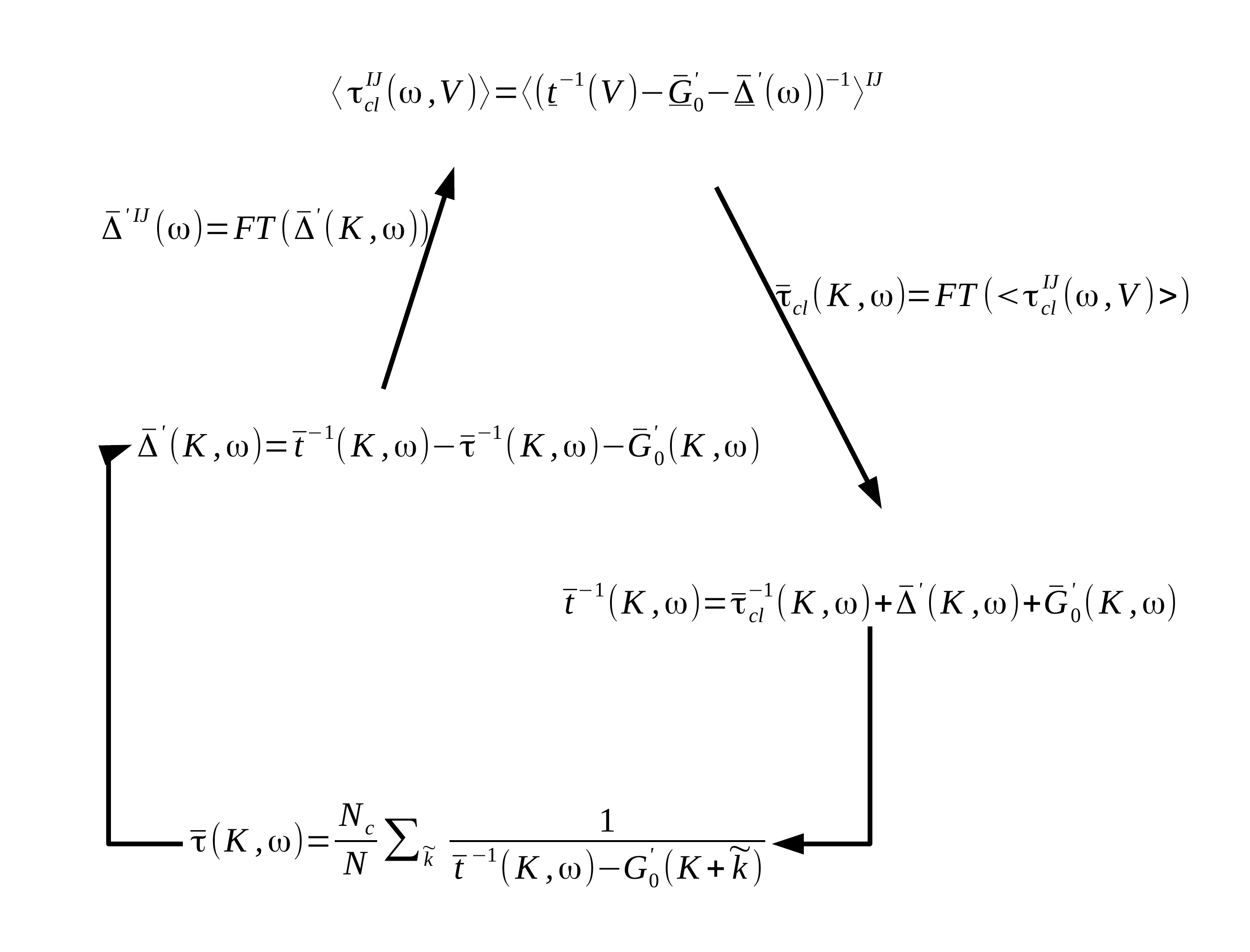}
\caption{The self-consistency loop of the KKR-NLCPA algorithm I  presented in Subsection~\ref{subsubsec:kkr1}.}
\label{fig:kkr1_loop-1}
\end{figure}

The self-consistency loop used in this ``KKR-NLCPA algorithm I" is illustrated in Fig.~\ref{fig:kkr1_loop-1} and the iterative procedures are described as follows:
\begin{itemize}
\item[1)] First we make a guess for the unknown cluster $\bar{t}(\K,\omega)$ scattering t-matrix (usually set to zero) and calculate the lattice coarse-grained scattering path operator $\bar{\tau}(\K,\omega)$ using Eq.~(\ref{eq: coarse-grained tau}).
\item[2)] Then we construct the effective-medium cluster renormalized interactor function $\bar\Delta^{'}(\K,\omega)$ by solving Eq.~(\ref{eq:-tau}) and Eq.~(\ref{eq:-tcl}), i.e., $\bar\Delta^{'}(\K,\omega)=\bar{t}^{-1}(\K,\omega)-\bar{\tau}(\K,\omega)^{-1}-\bar{G}_0^{'}(\K,\omega)$. 
\item[3)] Since the cluster problem is solved in real space, we Fourier transform the obtained cluster interactor $\bar{\Delta}^{'}(\K,\omega)$ to real space. We then solve the cluster problem and calculate the disorder-average scattering path operator $\langle {\underline\tau}_{cl}(\omega,V)\rangle=\left\langle \left(\underline{t}(V)^{-1}-\underline{\bar{G}}_0^{'}-\underline{\bar{\Delta }}^{'}(\omega)\right )^{-1}\right\rangle$.
\item[4)] Once the cluster problem is solved, we Fourier transform the obtained disorder-averaged path operator to the cluster reciprocal space with $\bar{\tau}_{cl}(\K,\omega)=FT(\langle\tau_{cl}(\omega,V)^{IJ}\rangle)$ and use it to construct new cluster $\bar{t}(\K,\omega)$ scattering matrix, i.e.,
$\bar{t}(\K,\omega)=\bar{\tau}_{cl}^{-1}(\K,\omega)+\bar{\Delta}^{'}(\K,\omega)+\bar{G}_0^{'}(\K,\omega)$.
\item[5)] We repeat steps 1-4 of the self-consistent procedure until $\bar{t}(\K,\omega)$ converges to the desired accuracy.
\end{itemize}

\subsubsection{ KKR-NLCPA algorithm II:\\ Local and non-local $t$-matrix  contributions}

The KKR-NLCPA algorithm of Refs.~\onlinecite{Biava,Rowlands_2003} with explicit separation of local and non-local contributions to the t-matrix is an alternative way of constructing the self-consistency. Here we show that this formalism can be derived from the one presented in the previous subsection ``KKR-NLCPA algorithm I". 

As discussed above, to explicitly distinguish between the local and non-local contributions to the scattering $t$-matrix, we define a quantity 
\begin{equation}
\bar{\alpha}(\K,\omega)=\bar t_l^{-1}(\omega)-\bar t ^{-1}(\K, \omega).
\label{eq:alpha2}
\end{equation}
Here $\bar{t}_{l}(\omega)$ is the local component of the scattering $t$-matrix, i.e., the single-site scattering matrix (we use a subscript $l$ to emphasize that $ \bar{t}_{l}(\omega)$ is a local quantity). Although $\bar{\alpha}(\K,\omega)$ describes the non-local contributions to the scattering $t$-matrix (in the same way as $\tilde {\alpha} (\K,\omega)$ describes non-local corrections to the DCA self-energy), in the KKR-NLCPA literature 
it is called the correction to the free propagator~\cite{Rowlands_2003}. Also note, that in Ref.~\onlinecite{Biava} this quantity is denoted as $\delta G(\K,\omega)$. 
Using Eq.~(\ref{eq:alpha2}) and making the DCA approximation with $\bar{t}(\k,\omega)=\bar{t}(\K,\omega)$, we rewrite the effective-medium lattice scattering path operator of Eq.~(\ref{eq. tau_lattice}) as
\begin{equation}
\bar{\tau}(\k,\omega)=\frac{1}{\bar{t}^{-1}_l(\omega)-\bar{\alpha}(\K,\omega)-G_0^{'}(\k,\omega)}.
\label{eq:tau_lattice1}
\end{equation}
with the lattice vector $\k=\K+\tilde{\k}$.

To find the yet unknown $\bar{t}^{-1}_l(\omega)$ and $\bar{\alpha}(\K,\omega)$, we consider the cluster path operator $\bar{\tau}_{cl}(\K,\omega)$ of Eq.~(\ref{eq:-tau}), with
\begin{equation}
\tau_{cl}^{-1}(\K,\omega)=\bar{t}_l^{-1}(\omega)-\bar{\alpha}(\K,\omega)-\bar{G}_0^{'}(\K,\omega)-\bar{\Delta}^{'}(\K,\omega),
\label{eq:taucl_nonlocal}
\end{equation}
where we used the fact that Eq.~(\ref{eq:-tcl}) can be rewritten as
\begin{equation}
t_{cl}^{-1}(\K,\omega)=\bar{t}_l^{-1}(\omega)-\bar{\alpha}(\K,\omega)-\bar{G}_0^{'}(\K,\omega)\,.
\label{tcl_nonlocal}
\end{equation}
This corresponds to Eq.(5) of Ref.~\onlinecite{Rowlands_2003}.

Next, to determine the effective-medium quantities we solve the cluster, with the disorder placed on each cluster site, embedded in the effective medium. Replacing $\bar{t}_l^{-1}(\omega)-\bar{\alpha}(\K,\omega)$ in Eq.~(\ref{eq:taucl_nonlocal}) with the disorder-dependent $t^{-1}(V)$ in Eq.~(\ref{eq:taucl_nonlocal}), we again obtain Eq.~(\ref{eq:-10-1-1-1}).

In the next step, we impose a self-consistency condition, which requires that placing a cluster in the effective medium on average does not produce additional scattering, i.e.,
\begin{equation}
\left<\tau_{cl}(\omega,V)\right>^{I J}=\bar\tau^{I J}_{cl}(\omega).\label{eq:selfcon_kkr2}
\end{equation}

To complete the self-consistency loop, we coarse-grain the scattering path operator of Eq.~(\ref{eq:tau_lattice1})
\begin{equation}
\bar{\tau}(\K,\omega)=\frac{N_c}{N}\sum_{\tilde{\k}} \frac{1}{\bar{t}^{-1}_l(\omega)-\bar{\alpha}(\K,\omega)-G_0^{'}(\k,\omega)}, 
\label{eq:-taucl-cg}
\end{equation}
where, $t^{-1}_l(\omega)-\bar{\alpha}(\K,\omega)=FT(\left<\tau_{cl}(\omega,V)\right>^{I J})^{-1}+G_0^{'}(\K,\omega)+\bar{\Delta}^{'}(\K,\omega)$,
according to Eq.~(\ref{eq:taucl_nonlocal}) and Eq.~(\ref{eq:selfcon_kkr2}).

After self-consistency is achieved, we can calculate the cluster coarse-grained
Green's function using Eq.~(\ref{eq:-3-2}):
\begin{eqnarray}
\bar{G}(\K,\omega)&=& \frac{N_c}{N}\sum_{\tilde{\k}}
 ( G_0(\K+\tilde{\k},\omega) \nonumber \\
&+& G_0(\K+\tilde{\k},\omega)\bar{\tau}(\K+\tilde{\k},\omega)G_0(\K+\tilde{\k},\omega)
),
\label{eq:-3-2}
\end{eqnarray}
where $G_0(\K+\tilde{\k},\omega)$ the bare lattice Green's function $ G_{0}(\K+\tilde{\k})=(\omega-{\epsilon}_{\K+\tilde{\k}}+i\eta)^{-1}$ and 
$\bar{\tau}(\K+\tilde{\k},\omega) = {\bar t_l(\omega)}^{-1}-\bar \alpha(\K,\omega)- G_{0}(\K+\tilde{\k},\omega)$. In the tight-binding Hamiltonian, the bare Green's function $G_0$ and the structure constant $G_0^{'}$ are identical. Nevertheless we still use separate symbols to distinguish the physical meaning of the quantities.	

\begin{figure}[h!]
\includegraphics[trim = 0mm 0mm 0mm 0mm,width=1\columnwidth,clip=true]{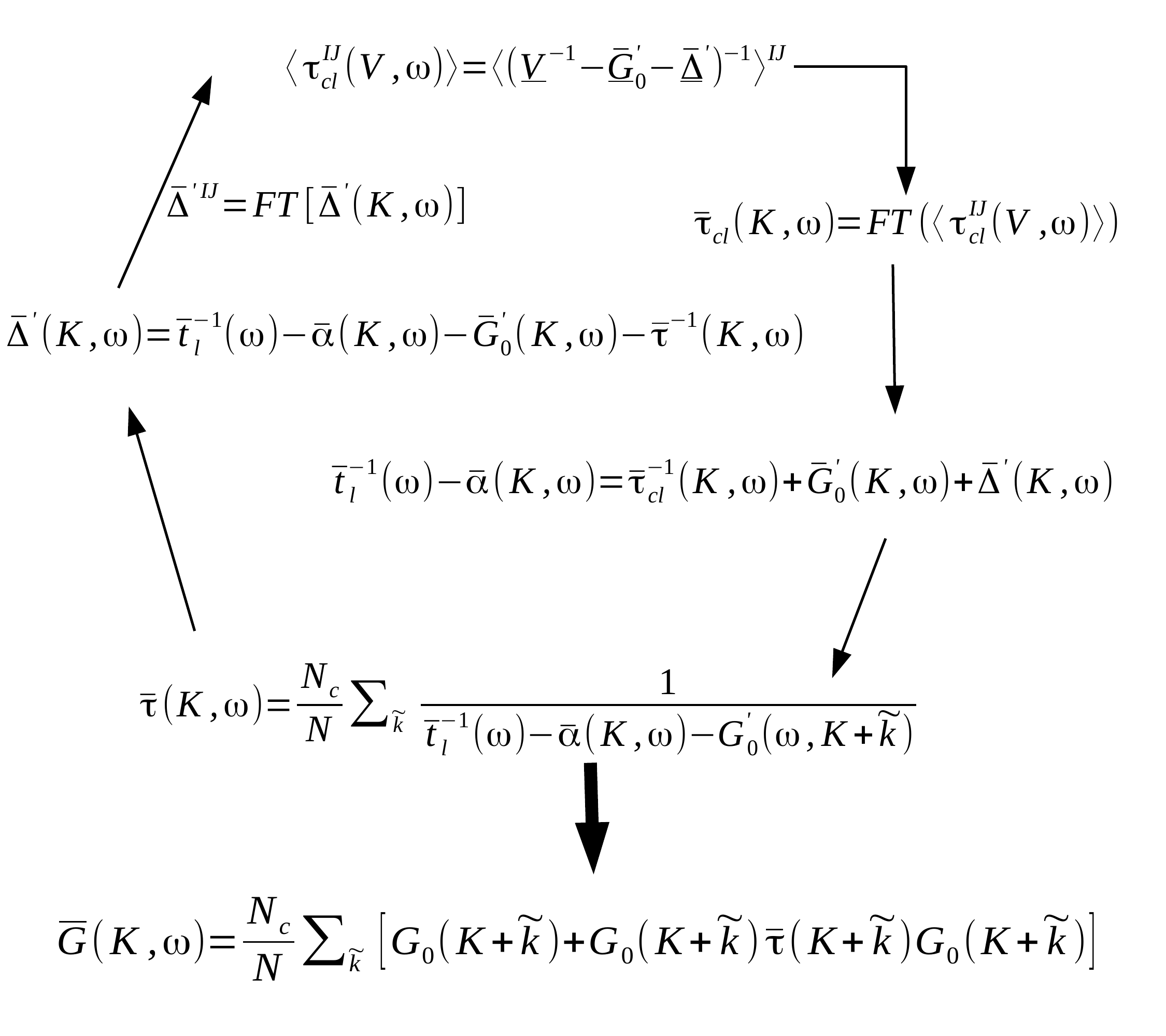}
\caption{The KKR-NLCPA algorithm II self-consistency loop applied to the tight binding model.}
\label{fig:kkr-scl}
\end{figure}

The self-consistency loop for the KKR-NLCPA algorithm II is presented in Fig.~\ref{fig:kkr-scl}. It is easy to see that, using $\bar{t}_l^{-1}(\omega)-\bar{\alpha}(\K,\omega)=\bar{t}^{-1}(\K,\omega)$, both KKR-NLCPA algorithm I and KKR-NLCPA algorithm II are equivalent.

\subsection{Relation between DCA and KKR-NLCPA quantities}
\label{sec:rel_KKR-DCA}
The application of the DCA and the KKR-NLCPA formalism to the tight-binding model discussed in Sec.~\ref{sec:DCA} and Sec.~\ref{sec:KKR-tau}, shows that there exist formal analogies between quantities and self-consistency equations used in these two approaches. For example, the effective cluster locator of the DCA approach, $g_{cl}(K,\omega)$, plays the role of the cluster $t$-matrix $t_{cl}(K,\omega)$ in the KKR-NLCPA formalism; the coarse-grained hopping matrix $\bar{W}^{IJ} $ is replaced by the free-space structure constant $\bar{G}_0'^{IJ}$; and the non-local contributions to the self-energy $\tilde{\alpha}(K,\omega)$ correspond to the non-local contribution to the scattering $t$-matrix  $\bar{\alpha} (K,\omega)$.  For completeness, in Table.~\ref{fig:DCA_KKR-Table}, we summarize the one-to-one correspondence between the DCA and the KKR-NLCPA equations.

In the following we explicitly show that quantities in the DCA are related to those in the KKR-NLCPA. In doing so we establish the formal equivalence of the two methods when applied to a tight-binding model. Using the obtained relationship, we construct an alternative Multiple-Scattering DCA (MS-DCA) algorithm which is a Green's function based multiple-scattering algorithm which allows one to calculate the disorder-averaged Green's function instead of the scattering path operator in self-consistent KKR-NLCPA loop. This step is necessary for the further implementation of the typical medium analysis in the multiple-scattering formalism. 

\subsubsection{DCA hybridization function and the NLCPA cluster renormalized interactor}
First, we show the relationship between the DCA hybridization function $\bar{\Delta}(\K,\omega)$ and the KKR-NLCPA renormalized interactor $\bar{\Delta}^{'}(\K,\omega)$. To establish this, we calculate the disorder-averaged Green's function of the cluster using the scattering-path operator
\begin{eqnarray}
\left<\underline{G}_{cl}(\omega)\right>&
=&
\left<\left(\underline{\mathcal{G}}^{-1}(\omega)-\underline{V}\right)^{-1}\right>\\ \nonumber
&=& \underline{\mathcal{G}}(\omega)+\underline{\mathcal{G}}(\omega) \left<\underline{\tau}_{cl}(\omega,V)\right> \underline{\mathcal{G}}(\omega)\,.
\label{eq:-gav}
\end{eqnarray}
Here the cluster scattering path operator matrix is given by
\begin{equation}
\underline{\tau}_{cl}(\omega,V)=\left( \underline{V}^{-1}-\underline{\mathcal{G}}(\omega) \right)^{-1}\,.
\label{eq:tau}
\end{equation}
Further, recalling Eq.~(\ref{eq:-10-1-1-1}) of the NLCPA procedure, the cluster path operator is 
\begin{equation}
\underline{\tau}_{cl}(\omega,V) =\left(\underline{t}^{-1}(V)
-\underline{\bar G}_{0}^{'}(\omega)-\underline{\bar{\Delta}}^{'} (\omega)  \right)^{-1}\,.
\label{eq:tau-kkr}
\end{equation}
Comparing Eqs.~(\ref{eq:tau}) and (\ref{eq:tau-kkr}), we find that
the cluster-excluded Green's function $\underline{\mathcal{G}}(\omega)$ satisfies the relation
\begin{equation}
\underline{\mathcal{G}} (\omega) =\underline{\bar G}^{'}_{0}(R)+\underline{\bar{\Delta}}^{'}(\omega)\,.
\label{eq: tau-gsc}
\end{equation} 

Next, in order to find the relation between the DCA hybridization function and the KKR-NLCPA interactor, we employ the expression for the cluster-excluded Green's function 
$\underline{{\mathcal{G}}}^{-1}(\omega)=\left(\omega \mathbb{I} - \underline{\bar W}-\underline{\bar \Delta}(\omega)\right)$. Thereby, the NLCPA cluster renormalized interactor $\bar{\Delta}^{'}$ is found to be related to the DCA hybridization function $\bar{\Delta}$ as
\begin{equation}
\underline{\bar{\Delta}}^{'}(\omega)=(\omega 
\mathbb{I} -\underline{\bar W}-\underline{\bar{\Delta}}(\omega))^{-1}-\underline{\bar G}_{0}^{'}\,.
\label{eq: del-del}
\end{equation}

Using this relationship, we now obtain an expression for the cluster Green's function  which can be calculated directly in the KKR-NLCPA self-consistency scheme, i.e.,
\begin{eqnarray}
\left<\underline{G_{cl}}(\omega)\right>&=&\left<\left(\underline{\mathcal{G}}^{-1}-\underline{V}\right)^{-1}\right>\\ \nonumber
                 &=& \left<\left(\left (\underline{\bar{\Delta}}^{'}(\omega)+\underline{\bar G}_{0}^{'}(\omega)\right)^{-1}-\underline{V}\right)^{-1}\right>\,.
\label{eq:-gimp-in-tau}
\end{eqnarray}

\begin{figure}[t!]
\includegraphics[trim = 0mm 0mm 0mm 0mm,width=1\columnwidth,clip=true]{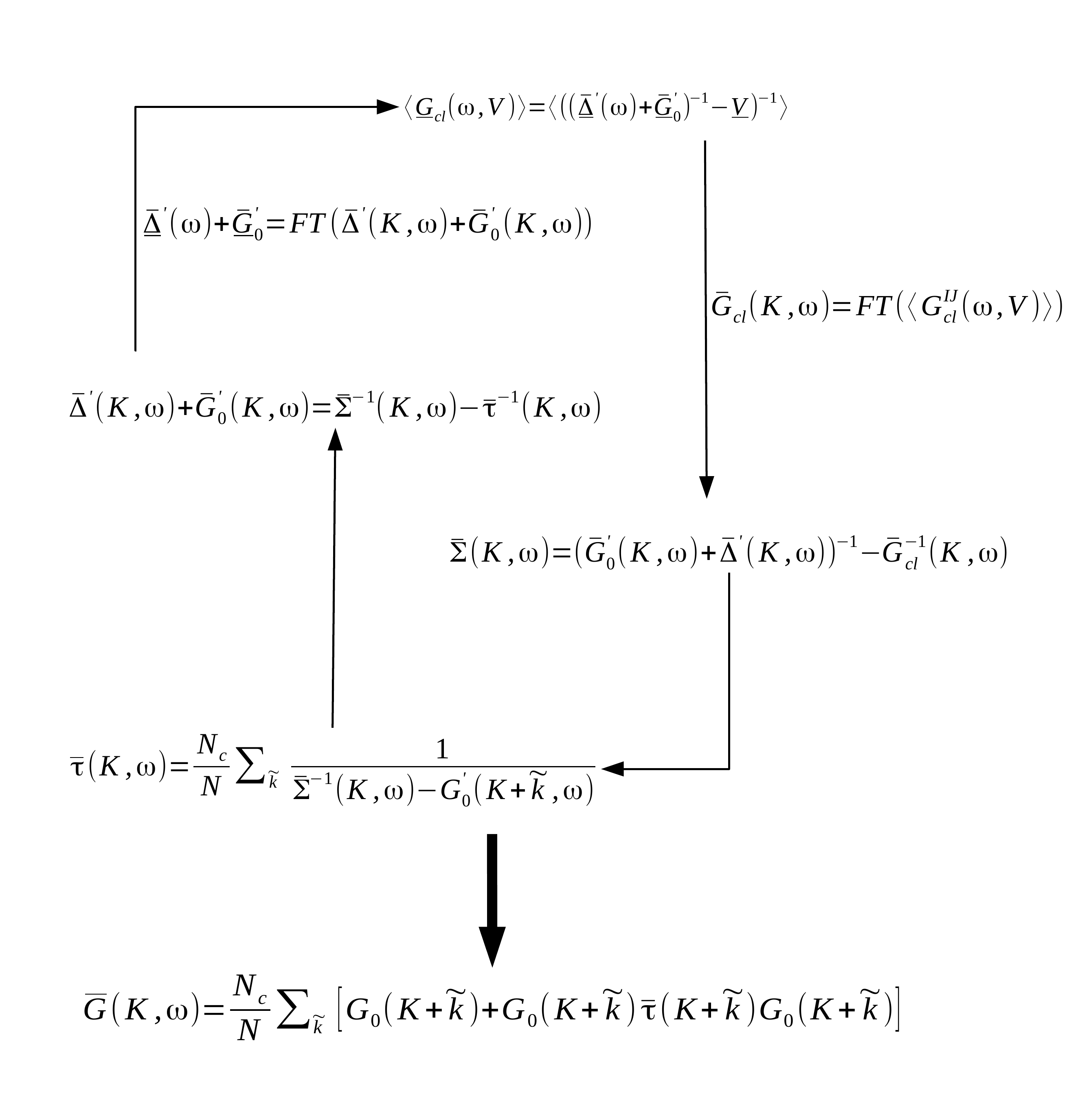}
\caption{Multiple-Scattering DCA: a Green's function based multiple scattering algorithm where we calculate the disorder-averaged Green's function instead of the scattering-path operator.}
\label{fig:KKR-GF}
\end{figure}

\subsubsection{DCA self-energy $\Sigma(\K,\omega)$ and the NLCPA 
effective corrections $\bar\alpha(\K,\omega)$}

Here we establish the relationship between the DCA self-energy and the non-local $t$-matrix corrections $\bar\alpha(\K,\omega)$ of the KKR-NLCPA.  In the latter, the lattice Green's function is given by
\begin{subequations}
\begin{eqnarray}
G(\k,\omega)&=&G_0(\k,\omega)+G_0(\k,\omega)\bar{\tau}(\k,\omega)G_0(\k,\omega) \\ \nonumber
&=&G_0(\k,\omega)+G_0(\k,\omega) \times \\ \nonumber
& & \frac{1}{\bar{t}_l^{-1}(\omega)-\bar{\alpha}(\K,\omega)-G_0(\k,\omega)}G_0(\k,\omega) \,.
\label{eq: tau-GF}
\end{eqnarray}
which can also be written as 
\begin{equation}
G(\k,\omega)=\frac{1}{G_0^{-1}(\k,\omega)-\frac{1}{\bar{t}_l^{-1}(\omega)-\bar{\alpha}(\K,\omega)}}\,.
\label{eq: tau-GF2}
\end{equation}
\end{subequations}
At the same time, in the DCA scheme the lattice Green's function is given by
\begin{equation}
G(\k,\omega)=\frac{1}{G_0^{-1}(k,\omega)-\bar{\Sigma}(\K,\omega)}\,.
\label{eq: DCA-GF}
\end{equation}
Comparing Eqs.~(\ref{eq: tau-GF2}) and (\ref{eq: DCA-GF}), we find
\begin{equation}
\bar{\Sigma}^{-1}(\K,\omega) =\bar{t}_l^{-1}(\omega) - \bar{\alpha}(\K,\omega)\,.
\label{eq: sigma-dG}
\end{equation}
This relationship shows how to obtain the self-energy in the KKR-NLCPA analysis.

\subsubsection{Multiple-Scattering DCA algorithms}
Using the obtained relationships between the DCA and KKR-NLCPA for the hybridization function and the self-energy, we now construct an alternative MS-DCA self-consistency loop where the disorder-averaged cluster Green's function instead of the scattering-path operator is calculated directly in the self-consistency. The diagram with the MS-DCA algorithm is shown in Fig ~\ref{fig:KKR-GF}, where, in addition to Eq.~(\ref{eq: del-del}),  Eq.~(\ref{eq:-gimp-in-tau}) and Eq.~(\ref{eq: sigma-dG}),  we also use the Dyson's equation with 
\begin{eqnarray}
\Sigma(\K,\omega)&=&\mathcal{G}^{-1}(\K,\omega)-G_{cl}^{-1}(\K,\omega)\nonumber \\
&=&\left(\bar{\Delta}^{'}(\K,\omega)+\bar{G}_0^{'}(\K,\omega)\right)^{-1}-G_{cl}^{-1}(\K,\omega)
\end{eqnarray}
As shown in Sec.~\ref{sec:KKR-TMDCA} below, this algorithm may be adapted to a typical medium approach.

\begin{widetext}

\begin{table}[!ht]
\caption{The correspondence between quantities appearing in the DCA and the KKR-NLCPA formalism.}
\label{fig:DCA_KKR-Table}
\begin{center}
\begin{tabular}{c|c}
\hline\\
DCA($N_c>1$) & KKR-NLCPA ($N_c>1$) \\
\hline \hline
Effective medium cluster Green's function & Effective medium cluster scattering-path operator \\
$\bar{G}_{cl}(\K,\omega) = \bar{g}_{cl}(\K,\omega) + \bar{g}_{cl}(\K,\omega) \bar{\Delta}(\K,\omega) \bar{G}_{cl}(\K,\omega)  $ &
$\bar{\tau}_{cl}(\K,\omega) = \bar{t}_{cl}(\K,\omega)+\bar{t}_{cl}(\K,\omega) \bar{\Delta}^{'}(\K,\omega)\bar{\tau}_{cl}(\K,\omega)$ \\ \hline
Isolated cluster Green's function in $K$-space & Isolated cluster scattering $t$-matrix \\
$ \bar{g}_{cl}(\K,\omega) = \frac{1}{\omega - \bar{\bar{\Sigma}}(\K,\omega)-\bar{\epsilon}(\K)} $ &
$ \bar{t}^{-1}_{cl}(\K,\omega) = \frac{1}{\bar{t}^{-1}(\K,\omega)-\bar{G}_0^{'} (\K,\omega)}  $    \\ \hline
Effective medium cluster Green's function & Effective medium cluster path operator \\ 
with separate local and non-local contributions & with separate local and non-local contributions \\
\begin{tabular}{c}
$\bar{G}_{cl}^{-1}(\K,\omega) = \bar{g}_{cl}^{-1}(\K,\omega) -\bar{\Delta(\K,\omega)} $ 
\\
$\bar{g}_{cl}^{-1}(\K,\omega) = g_l^{-1}(\K,\omega)-\tilde{\alpha}(\K,\omega)-\bar{\epsilon}(\K)$
\end{tabular} &
\begin{tabular}{c}
$ \bar{\tau}_{cl}^{-1}(\K,\omega) = \bar{t}_{cl}^{-1}(\K,\omega) -\bar{\Delta}^{'}(\K,\omega)$  
\\
$t_{cl}^{-1}(\K,\omega)=t^{-1}_l(\omega)-\bar{\alpha}(\K,\omega)-\bar{G}_0^{'}(\K,\omega)$
\end{tabular}
 \\ \hline
Non-local contributions of the self-energy & Non-local contributions of the scattering $t$-matrix \\
$\tilde{\alpha}(\K,\omega)=\bar {\Sigma}(\K,\omega)-\bar{\Sigma}_l(\omega)$ 
& 
$ \bar{\alpha}(\K,\omega)=\bar t_l^{-1}(\omega)-\bar t ^{-1}(\K, \omega)  $
\\ \hline
\end{tabular}
\end{center}
\end{table}
\end{widetext}

\section{Typical medium Multiple-Scattering formalism}
\label{sec:KKR-TMDCA}
While the DCA incorporates spatial correlations which are missing in the CPA, the {\it average} DOS calculated in the DCA~\cite{PhysRevB.89.081107} is not critical at the Anderson  transition,~\cite{Thouless1970,Thouless} and hence cannot be used as an order parameter. To identify Anderson localized states, one has to calculate the {\it typical} density of states (TDOS). Indeed, it has been demonstrated that the TDOS vanishes for localized states, and hence, can be used as a proper order parameter (Refs.~\onlinecite{Vlad2003} and~\onlinecite{PhysRevB.89.081107}).

In the typical medium theory the self-consistency involves the TDOS which vanishes continuously as the strength of the disorder increases towards the critical point (Refs.~\onlinecite{Vlad2003} and~\onlinecite{PhysRevB.89.081107}). As a consequence, there is an additional step needed to calculate this order parameter. 

In the following, we show how to apply the typical medium analysis to the tight-binding model within the multiple scattering approach (MS-TMDCA). We have already shown in Fig.~\ref{fig:KKR-GF} how to calculate the average Green's function in the self-consistency loop.  To incorporate the typical medium analysis in the KKR-NLCPA, we modify the effective medium by replacing the algebraically averaged Green's function by its typical, i.e., geometrically averaged, counterpart. Thereby translational invariance is restored by the disorder average everywhere in the distribution of the density of states, including the average and typical values. To construct the typical Green's function, we use the same ansatz as in Ref.~\onlinecite{PhysRevB.89.081107}). For each cluster configuration, we obtain the cluster density of states diagonal in the wave number,
$\rho_{cl}(\K,\omega,V)=-\Im G_{cl} (\K,\K,\omega,V)/\pi$, assuming that the off-diagonal contributions vanish.  We also calculate the LDOS on the cluster $\rho_{cl}^{II}(\omega,V)=-\Im G_{cl}^{II} (\omega,V)/\pi$.  
Then, we calculate the typical (geometrically averaged) density of states as 
\begin{eqnarray}\label{eq:ansatz1a}
\bar{\rho}_{typ}(\K,\omega) &=& \exp\left(\frac{1}{N_c} \sum_{I=1}^{N_c} \left\langle \ln \rho_{cl}^{II} (\omega,V)\right\rangle\right) \nonumber\\
&\times & \left\langle \frac{\rho_{cl}(\K,\omega,V)}{\frac{1}{N_c} \sum_{I} \rho_{cl}^{II} (\omega,V)} \right\rangle \,.
\label{eq:rho_typ}
\end{eqnarray}
Here, as in the TMDCA, the local part of the cluster-momentum-resolved typical density of states is separated and treated with geometrical averaging over the disorder configurations, to avoid self-averaging as the cluster size increases. The form of the proposed typical density of states of Eq.~(\ref{eq:rho_typ}) recovers the local TMT at $N_c=1$ limit and reduces to the DCA scheme at weak disorder strength. 

\begin{figure}[htb!]
\includegraphics[trim = 0mm 0mm 0mm 0mm,width=1\columnwidth,clip=true]{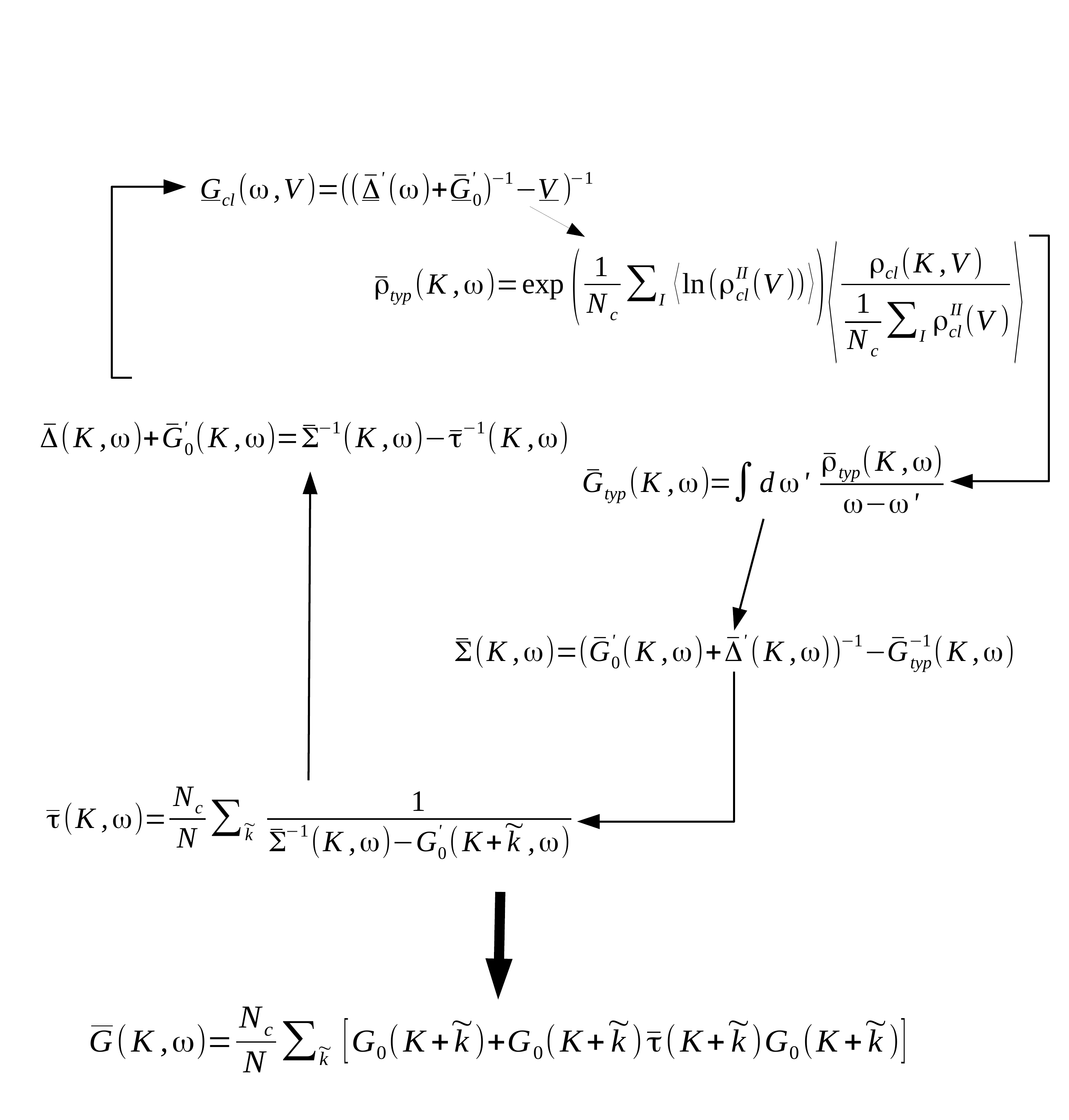}
\caption{MS-TMDCA: a Green's function based typical medium multiple scattering algorithm.}
\label{fig:KKR-Gtyp}
\end{figure}

By using the Hilbert transform, we obtain the typical Green's function from Eq.~(\ref{eq:ansatz1a}) as
\begin{equation}
\bar{G}_{typ}(\K,\omega)=\int d\omega' \frac{\bar{\rho}_{typ}(\K,\omega')}{\omega-\omega'}\,,
\label{eq:ansatz1b}
\end{equation}
which replaces the average Green's function in the self-consistency loop, Fig.~\ref{fig:KKR-GF}.   The resulting algorithm for the MS-TMDCA applied to the tight-binding model is shown in Fig.~\ref{fig:KKR-Gtyp}.
As discussed previously\cite{y_zhang_16}, the form of Eq.~(\ref{eq:ansatz1a}) is not unique. To be able to describe the transition it should contain the order parameter (the first term in the product), which vanishes at the transition.  A possible, but different form reads~\cite{y_zhang_16}
\begin{eqnarray}\label{eq:ansatz2}
\bar{G}_{typ}(\K,\omega) &=& \exp\left(\frac{1}{N_c} \sum_{I=1}^{N_c} \left\langle \ln \rho_{cl}^{II} (\omega,V)\right\rangle\right) \nonumber\\
 &\times& \left\langle \frac{G_{cl}(\K,\omega,V)}{\frac{1}{N_c} \sum_{I} \rho_{cl}^{II} (\omega,V)} \right\rangle\,. 
\end{eqnarray}
It dispenses with the need for the Hilbert transform so that results at different frequencies are independent by calculating the typical $\bar{G}_{typ}$ directly, at the expense of not recovering the TMT when $N_c=1$.  Nevertheless, it quickly converges to produce the same results as Eq.~(\ref{eq:ansatz1a}) and Eq.~(\ref{eq:ansatz1b}) for moderate cluster sizes, including $N_c=38$, but will not be further employed in this discussion.
It is important to note that the typical Green's function, $G_{typ}(\K,\omega)$, is used only to calculate the effective medium of the cluster problem.  This geometrically averaged reference system carries no physical meaning other than the order parameter. Since experimental measurements of the single-particle spectra (which also determine the Fermi level), and of transport, two-particle spectra, etc.\ involve averages over large regions, these experiments are described by arithmetically averaged Green's functions obtained from functional derivatives of the arithmetically averaged free energy.

In Tab.~\ref{table:acronymns} we summarize the naming conventions and acronyms for the different algorithms discussed in this manuscript.   Here, in each abbreviation, we use the letters TM to indicate a typical medium approach, the letters DCA for a Green's function based approach, and NLCPA a scattering path operator ($\tau$) based approach.  
\begin{table}
\begin{tabular}{p{1.4cm}|p{6.6cm}}
\bf{Acro.} & \bf{Definition} \\
\hline
DCA   & Dynamical Cluster Approximation, which calculates the average Green's function (Figs.~\ref{fig:DCA_loop} and \ref{fig:DCA_loop2})\\
\hline
KKR-NLCPA & A multiple scattering algorithm which calculates the average scattering path operator (Figs.~\ref{fig:kkr1_loop-1} and \ref{fig:kkr-scl})\\
\hline
MS-DCA & A multiple scattering algorithm, calculating the average Green's function (Fig.~\ref{fig:KKR-GF})\\
\hline
MS-TMDCA & A multiple scattering algorithm calculating the typical Green's function (Fig.~\ref{fig:KKR-Gtyp})\\
\hline
\end{tabular}
\caption{\label{table:acronymns}
Acronyms of the algorithms used.}
\vspace{-0,4cm}
\end{table} 

\section{Numerical results for the tight-binding model}
\label{sec:results} 

In this section, we will present the results obtained using the KKR-NLCPA formalism as compared with the standard DCA and the typical medium DCA using the tight-binding model of Sec.~\ref{sec:model}.

\subsection{Effective medium with arithmetic average: KKR-NLCPA vs.\ DCA}
As a starting point, we discuss results from benchmarking the KKR-NLCPA formalism with the DCA. 

As shown in Eq.~(\ref{eq: tau-gsc}) in Subsection~\ref{sec:rel_KKR-DCA}, the DCA cluster excluded Green's function is related to the KKR-NLCPA cluster renormalized interactor through 
$\mathcal{G}(\K,\omega)=\bar G_0^{'}(\K,\omega)+\bar{\Delta}^{'}(\K,\omega)$. 
We use this relationship in the construction of the multiple-scattering algorithms which employ the arithmetically averaged and typical Green's functions.
\begin{figure}[h!]
\includegraphics[trim = 0mm 0mm 0mm 0mm,width=1.\columnwidth,clip=true]{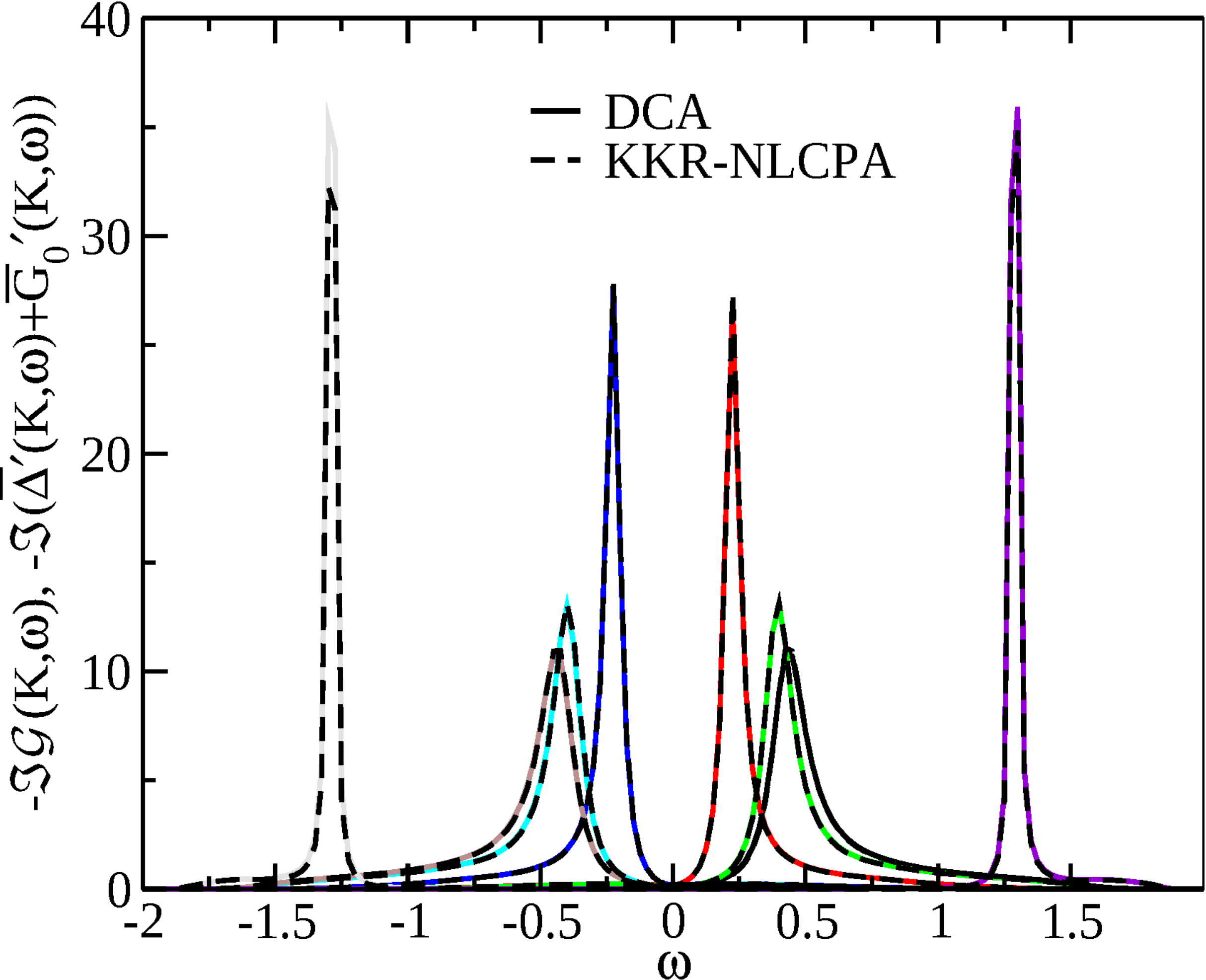}
\caption{Comparison of the imaginary parts of the cluster excluded Green's function $\mathcal{G}(\K,\omega)$  from the DCA procedure and $\bar G_0^{'}(\K,\omega)+\bar{\Delta}^{'}(\K,\omega)$ calculated in the KKR-NLCPA procedure for $N_c=38$ at disorder strength, V$_A=0.7$ and concentration, c$_A=0.5$. The solid lines depict the DCA results while the dash lines are the corresponding KKR-NLCPA and their associated momenta $\K$ correspond to each of the eight distinct cells obtained using the point-group symmetry of the cluster. }
\label{fig:gsc}
\end{figure} 

To demonstrate this numerically, we show in Fig.~\ref{fig:gsc} a comparison of 
the imaginary parts of $\mathcal{G}(\K,\omega)=\left(\omega-\bar{\epsilon}(\K)-\bar{\Delta}(\K,\omega)\right)^{-1}$ from the standard DCA and $\bar G_0^{'}(\K,\omega)+\bar{\Delta}^{'}(\K,\omega)$ calculated from the KKR-NLCPA algorithm shown in Fig.~\ref{fig:kkr-scl} for the binary alloy disorder configurations for the concentration $c_A=0.5$ at disorder strength $V_A=0.7$ for the cluster size $N_c=38$. In Fig.~\ref{fig:gsc}, the solid lines are the DCA results while the dash lines are the corresponding KKR-NLCPA. As evident from the plots, the two formalisms agree with each other within our numerical accuracy. This shows that the averaged medium KKR-NLCPA/DCA using the tight-binding model adequately reproduces the required behavior as in the DCA. 

\begin{figure}[h!]
\includegraphics[trim = 0mm 0mm 0mm 0mm,width=1\columnwidth,clip=true]{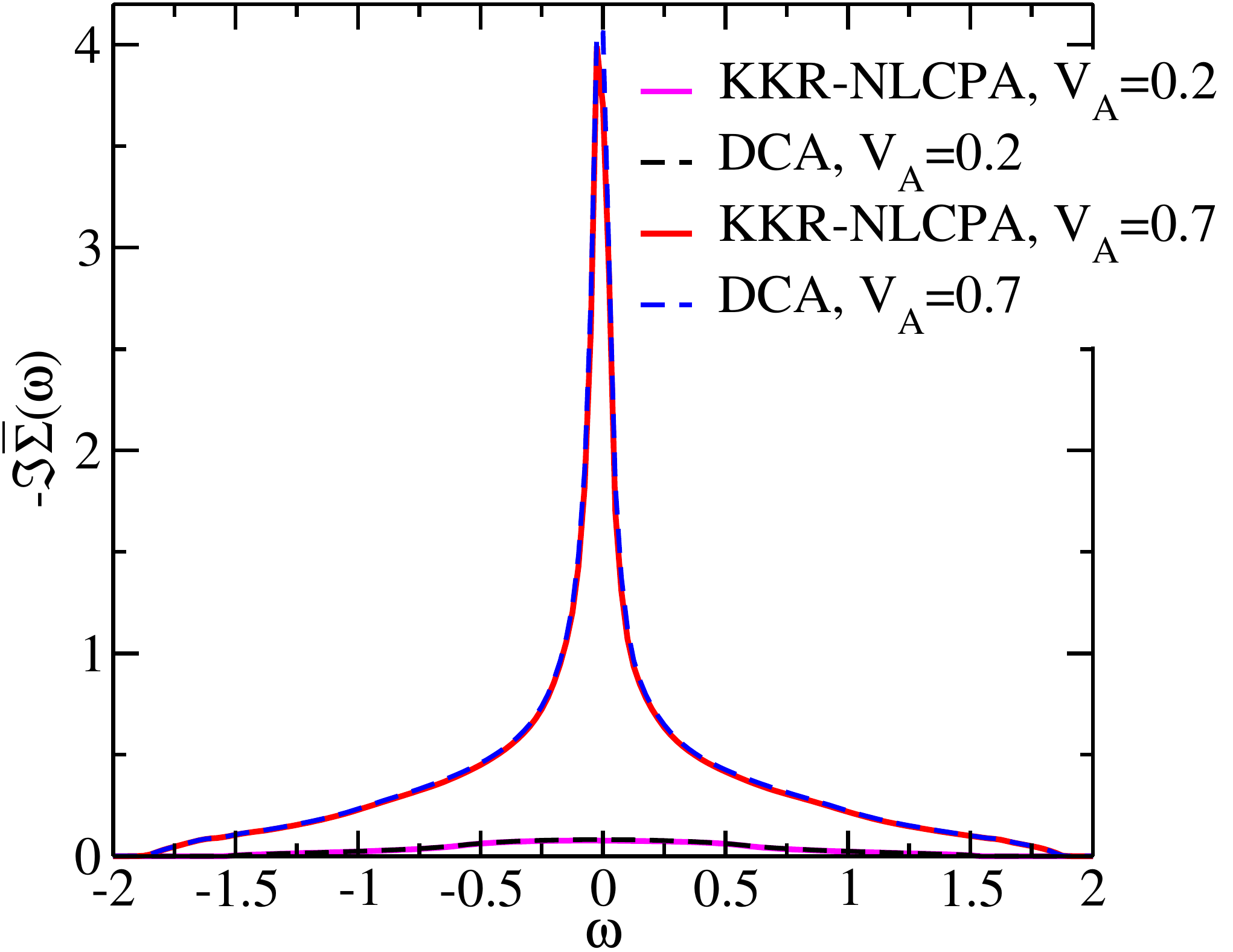}
\caption{A numerical comparison of the local self-energy $\frac{1}{N_c}\sum_\K\bar\Sigma(\K,\omega)$ obtained from the DCA procedure
 and the local KKR-NLCPA self-energy $\frac{1}{N_c}\sum_\K [t_l^{-1}(\omega)-\bar\alpha(\K)]^{-1}$. Here, solid lines are the DCA results and 
 dash lines are their KKR-NLCPA counterparts.}
 \label{fig:sigma-loc}
\end{figure}
\begin{figure}[h!]
\includegraphics[trim = 0mm 0mm 0mm 0mm,width=1\columnwidth,clip=true]{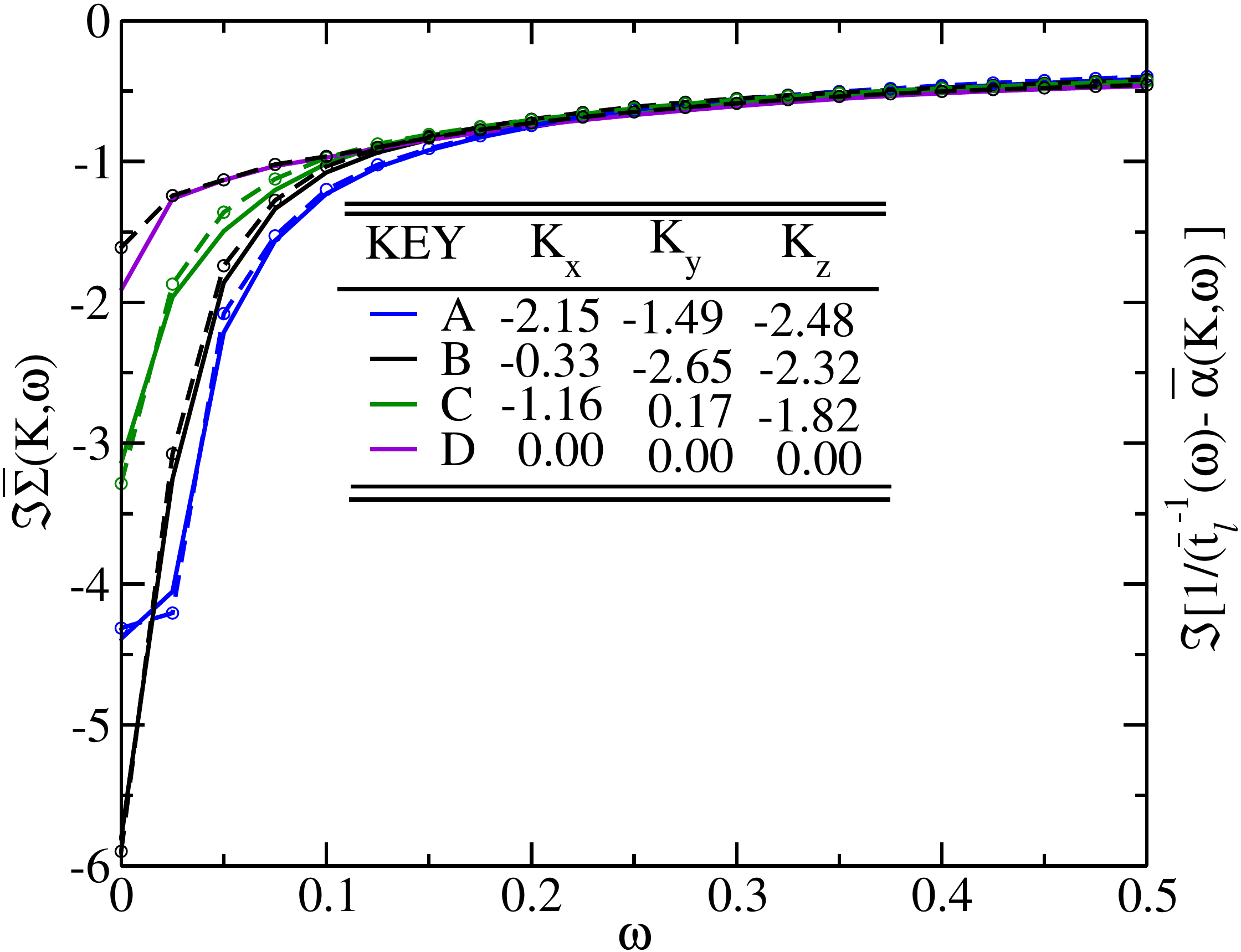}
\caption{A numerical comparison of the $\K$-resolved self-energy $\bar\Sigma(\K,\omega)$ obtained from 
the DCA procedure and the KKR-NLCPA self-energy $[\bar{t}_l^{-1}(\omega)-\bar\alpha(\K)]^{-1}$ at $V_A=0.7$ and $c_A=0.5$. 
We observe again that two procedure are numerically equivalent. Here, the solid lines are the DCA
results while the dashed lines depict their KKR-NLCPA counterparts. Note, the labels A--D and their
associated momenta $\K$ depict each of the four distinct cells obtained using the point-group and
particle-hole symmetry ($\bar{\rho}(\K,\omega) = \bar{\rho}(\Q-\K,-\omega)$, with $\Q= (\pi,\pi,\pi)$) of the
cluster. 
}
\label{fig:sigma-nonloc} 
\end{figure}

 
To perform a further numerical check of these expressions we compare the DCA self-energy $\bar\Sigma(\K,\omega)$ to the KKR-NLCPA (Fig.~\ref{fig:kkr-scl}) $[\bar{t}_l^{-1}(\omega)-\bar\alpha(\K,\omega)]^{-1}$. The results for the imaginary part of the local cluster average self-energy $\Im \bar{\Sigma}(\omega)=\frac{1}{N_c}\sum_{\K}\Im\bar{\Sigma}(\K,\omega)$ is shown in Fig.~\ref{fig:sigma-loc} and that for the imaginary part of the non-local momentum-resolved $\Im\bar{\Sigma}(\K,\omega)$ is shown in Fig.\ref{fig:sigma-nonloc}.  In Fig.~\ref{fig:sigma-nonloc}, the labels A--D and their associated momenta $\K$ correspond to each of the four distinct cells obtained using the point-group and particle-hole symmetry 
($\bar{\rho}(\K,\omega) = \bar{\rho}(\Q-\K,-\omega)$, with $\Q= (\pi,\pi,\pi)$) of the cluster. For the local self-energy, Fig.~\ref{fig:sigma-loc}, the two disorder strengths, $V_A=0.2$ and 0.7, correspond to the weak disorder and the band split regimes, respectively. 
In both cases, a good agreement between the two formalisms is observed. Similarly, as seen from Fig.~\ref{fig:sigma-nonloc}, we find a very good agreement between the non-local self-energies calculated within the DCA and KKR-NLCPA formalisms, respectively. Here we present results for the large disorder $V_A=0.7$ strength only, since at low disorder the non-local effects are negligible.

\begin{figure}[h!]
\includegraphics[trim = 0mm 0mm 0mm 0mm,width=1\columnwidth,clip=true]{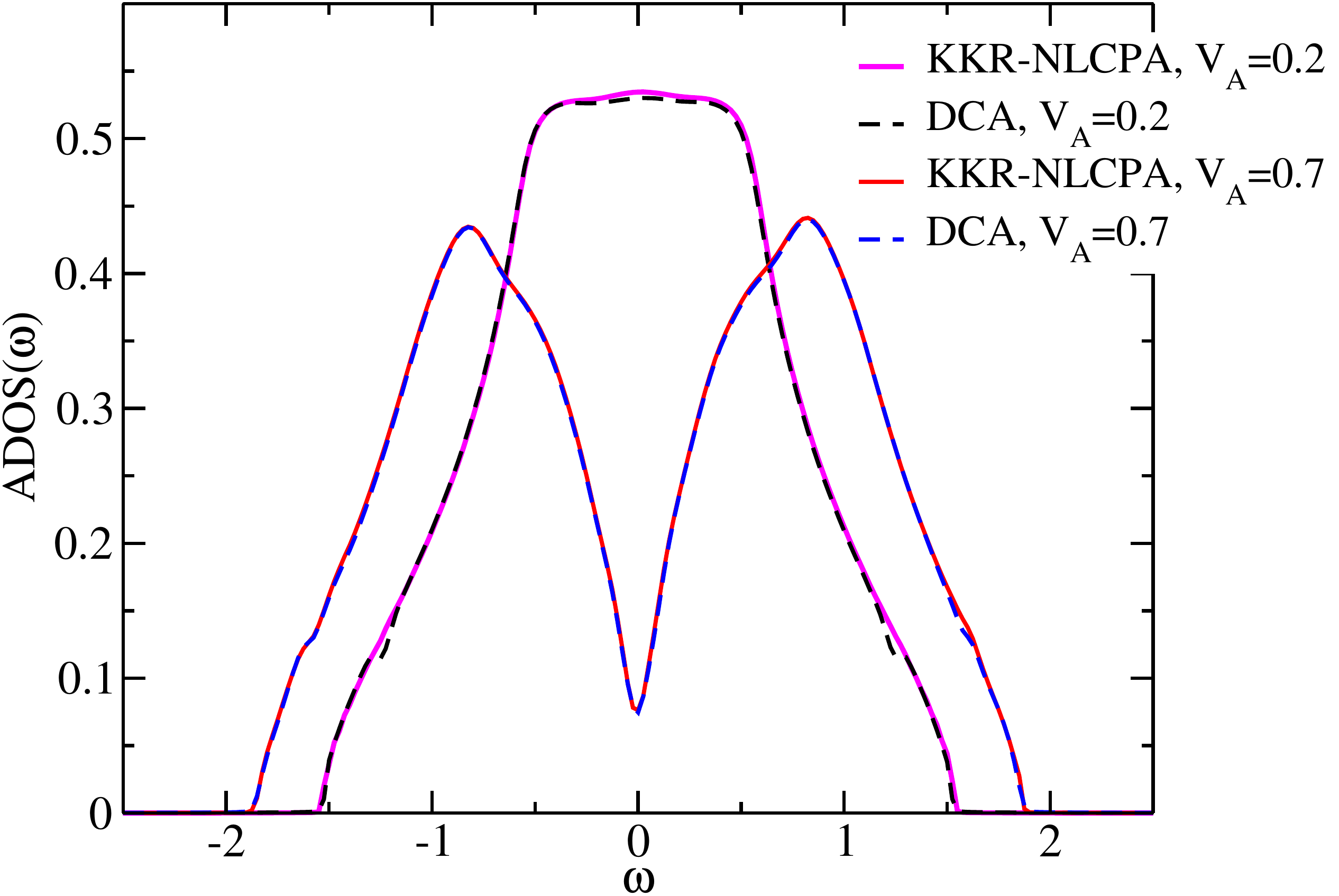}
\caption{The arithmetically averaged local density of states (ADOS) 
obtained from the DCA and KKR-NLCPA calculations. We present data for the binary alloy disorder with $c_A=0.5$, for
cluster size $N_c=38$ at small disorder $V_A=0.2$ and larger disorder $V_A=0.7$. 
The results  show that the two procedures are equivalent when applied to the tight-binding model.}
\label{fig:DoS}
\end{figure} 

To conclude our benchmarking of the multiple scattering KKR-NLCPA approach with the DCA, we show in Fig.~\ref{fig:DoS} a plot of the arithmetically averaged density of states (ADOS) obtained using the DCA algorithm I of Fig.~\ref{fig:DCA_loop} and the KKR-NLCPA of Fig.~\ref{fig:kkr-scl}.  This again is a good quantity to check the reliability of the developed KKR-NLCPA formalism within the tight-binding model. The data are for $c_A=0.5$ at weak disorder $V_A=0.2$ and strong disorder $V_A=0.7$ using a finite cluster $N_c=38$. The results clearly demonstrate that the two procedures are equivalent when applied to the tight-binding model.

\subsection{Typical medium: MS-TMDCA vs.\ TMDCA}
\begin{figure}[h!]
\includegraphics[trim = 0mm 0mm 0mm 0mm,width=1\columnwidth,clip=true]{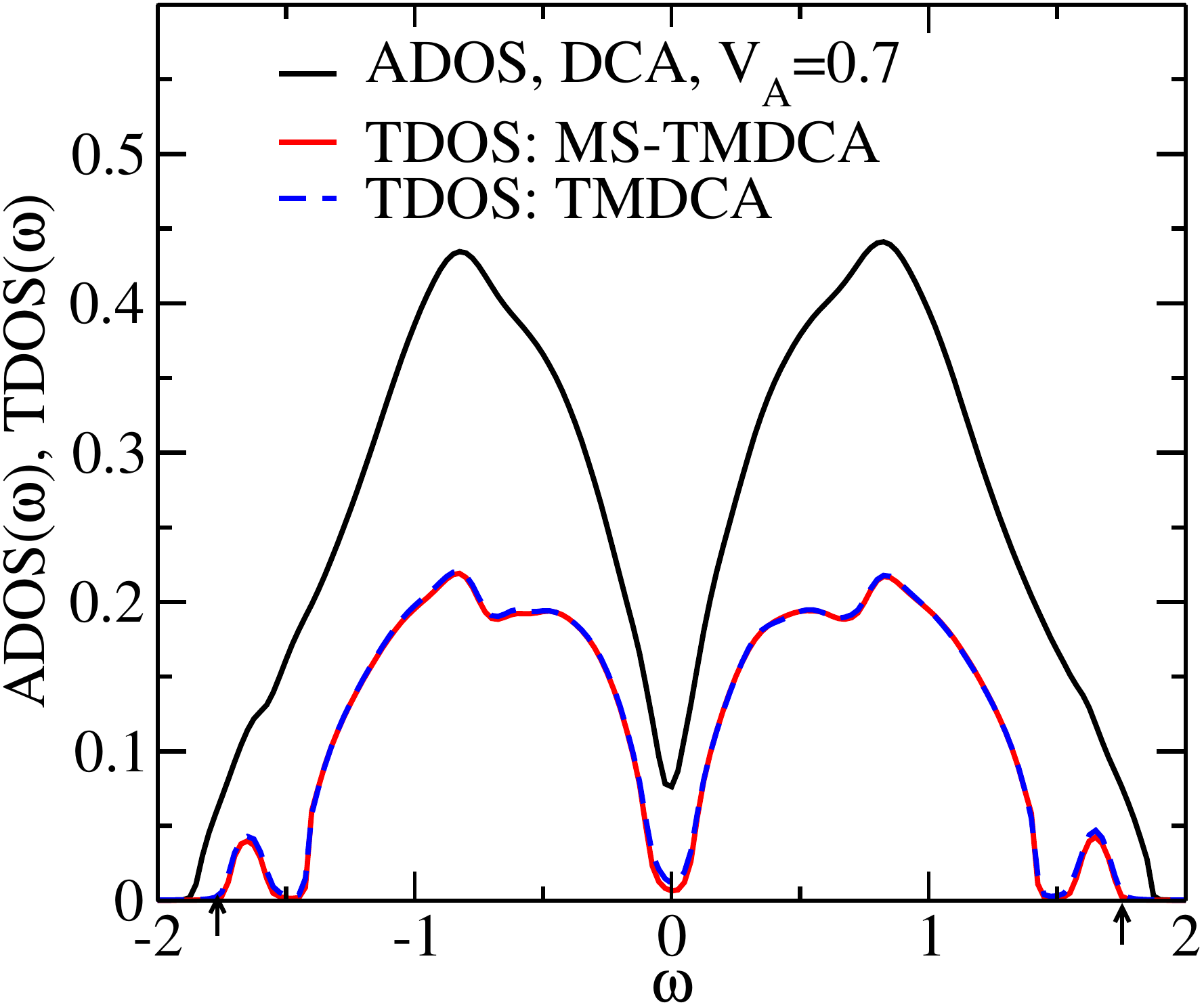}
\caption{ADOS calculated in DCA and KKR-NLCPA procedures and TDOS calculated in MS-TMDCA and TMDCA procedures. States within the mobility edge marked by arrows are extended.}
\label{fig:TDOS}
\end{figure}

In the typical medium analysis, the effective medium is characterized by a geometrically averaged typical density of states. To demonstrate that the proposed MS-TMDCA formalism captures correctly the Anderson localization transition, we compare our results of the typical density of states obtained from the MS-TMDCA procedure described in Fig.~\ref{fig:KKR-Gtyp} with the ones obtained using the TMDCA scheme~\cite{PhysRevB.89.081107}. As shown in Fig.~\ref{fig:TDOS} for a finite cluster size, $N_c=38$ at disorder strength, $V_A=0.7$
the ADOS from the standard DCA (of Fig.~\ref{fig:DCA_loop}) and the KKR-NLCPA (of Fig.~\ref{fig:kkr-scl}) are the same (black line). 

The results from the TMDCA of Ref.~\onlinecite{Ekuma2015} and the MS-TMDCA of Fig.~\ref{fig:KKR-Gtyp} are also nearly indistinguishable.  Here the TDOS is finite for the extended states (inside the mobility edge marked by arrows), and vanishes for the localized states at the top and bottom of the band and at the band center.  By contrast, the ADOS remains finite even for localized states.  With increasing $V_A$ it is possible to drive also the ADOS to zero at the band center due to the alloy disorder band splitting~\cite{Velicky}.

We note that the above formalism applies to the single-band case. However, realistic systems often have many bands. Thus, in electronic structure calculations based on the multiple-scattering typical-medium dynamical cluster approximation formalism the above quantities are matrices in angular momentum space. This will require some modifications to the proposed MS-TMDCA scheme. In particular, for the ansatz Eq.~(\ref{eq:ansatz1a})  for the multi-orbital TMDCA formalism one may perform a geometric average for the diagonal terms $l=m$ and a linear average for the off-diagonal $l\neq m$ components~\cite{y_zhang_15a}.  The second ansatz, Eq.~(\ref{eq:ansatz2}) requires less modification in that only a linear average of the Green function matrix occurs in the numerator, which appears to improve the numerical stability of the algorithm.

\section{Conclusion}
\label{sec:conclusion}

In this paper we detail the construction of the typical medium approach in the framework of the multiple-scattering formalism.  The constructed Multiple Scattering Typical Medium DCA (MS-TMDCA) formalism is a reformulation of the typical medium DCA~\cite{PhysRevB.89.081107} which has been successfully applied to model-Hamiltonian systems to quantitatively study and detect Anderson localization. 

Being motivated by the need for the development of appropriate numerical tools to study strong disorder effects in first principle calculations, we first provide a detailed comparison of two major effective medium algorithms used in the model Hamiltonian DCA~\cite{m_jarrell_01a} community and the real material multiple-scattering KKR-NLCPA community~\cite{Biava,Rowlands_2003,Rowlands_2006}. To provide a bridge between the DCA and multiple-scattering approaches, we demonstrate explicitly that these two approaches when applied to a tight-binding Hamiltonian are equivalent. We pay particular attention to the non-trivial relations between the key quantities in the DCA and KKR-NLCPA methods.  These relationship were used to construct various self-consistency procedures, including the MS-DCA which is a generalization of the DCA using the multiple-scattering language. 
Because the MS-DCA calculates the disorder-averaged Green's function rather than the scattering path operator used in the KKR-NLCPA, it can be readily generalized into a typical medium approach such as the MS-TMDCA (a multiple scattering algorithm).

As an application of the MS-TMDCA formalism developed here we solve a tight-binding model on a cubic lattice with $N_c=38$ sites for various values of binary disorder strength. To check the validity of our mapping of the TMDCA to the multiple scattering approach, we compare numerical results for the density of states and the self-energy calculated within the DCA-based and the multiple scattering-based approaches (both for the algebraically averaged and geometrically averaged components). We find that the results from both approaches are numerically the same, affirming the successful mapping of the DCA-based formalism to a MS based formalism.

The construction and application of the MS-TMDCA formalism to the tight-binding model presented in this paper is the first step towards an extension of the typical medium methodology into an ab-initio theory of disordered materials which is sensitive to Anderson localization.  The material-specific implementation of this approach will be our next step. 

\begin{acknowledgments}
We thank J.\ Moreno for useful conversations. This work (H.T., Y.Z., and M.J.) 	was supported by the National Science Foundation under the NSF EPSCoR Cooperative Agreement No.\ EPS-1003897 with additional support from the Louisiana Board of Regents. LC thanks L.\ Vitos for valuable discussions. LC and DV acknowledge funding by the Deutsche Forschungsgesellschaft (DFG) through the Transregional Collaborative Research center TRR80.
\end{acknowledgments}

\bibliographystyle{general}

\bibliography{TMDCA_KKR}

\end{document}